\documentclass[10pt,journal,compsoc]{IEEEtran}
\usepackage{longtable}
\usepackage{amssymb}
\usepackage{amsmath}
\usepackage{algorithm, algorithmic}
\usepackage{multirow}
\usepackage{setspace}
\usepackage{graphicx}
\usepackage{makecell}
\usepackage{tabularx}
\usepackage{caption}
\usepackage{subcaption}
\usepackage{hyperref}
\usepackage{pdflscape}
\usepackage{slashbox}
\usepackage{booktabs}
\newcommand{\tabitem}{~~\llap{\textbullet}~~}
\newcommand{\RNum}[1]{\uppercase\expandafter{\romannumeral #1\relax}}

\newcolumntype{S}{>{\centering\arraybackslash}m{1.5em}}

\newcommand{\figref}[1]{Figure~\ref{#1}}
\newcommand{\tblref}[1]{Table~\ref{tbl:#1}}

\usepackage{cite}
\usepackage{amssymb}
\usepackage{amsmath}
\usepackage{algorithm, algorithmic}
\usepackage{multirow}
\usepackage{setspace}
\usepackage{graphicx}
\usepackage[ampersand]{easylist}
\usepackage{caption}
\usepackage{subcaption}
\usepackage[table,xcdraw]{xcolor}
\usepackage{pdfpages}
\usepackage{adjustbox}
\usepackage[font=small,labelfont=bf]{caption}
\usepackage{cite}

\begin{document}
	\title{Personality Assessment from Text for Machine Commonsense Reasoning}

	\author{Niloofar~Hezarjaribi, Zhila~Esna Ashari, James~F.~Frenzel, Hassan~Ghasemzadeh,
		and~Saied~Hemati
		\thanks{This work was supported in part by a grant awarded by Samsung Global Research Outreach (GRO) program in 2016. Niloofar~Hezarjaribi, Zhila~Esna Ashari and  Hassan~Ghasemzadeh are with the School of Electrical Engineering and Computer Science, Washington State University, Pullman, WA 99164-2752, USA. Emails \{n.hezarjaribi, z.esnaashariesfahan, hassan.ghasemzadeh\}@wsu.edu. James~F.~Frenzel and Saied Hemati are with the Department of Electrical and Computer Engineering, University of Idaho, Moscow, ID, 83844-1023, USA. Emails: \{jfrenzel, shemati\}@uidaho.edu.}}
	\markboth{}%
	{N. Hezarjaribi, et al. ``Text-Based Personality Assessment for Machine Commonsense Reasoning''}

\IEEEtitleabstractindextext{%
\begin{abstract}
This article presents {\it PerSense}, a framework to estimate human personality traits based on expressed texts and to use them for commonsense reasoning analysis. The personality assessment approaches include an aggregated Probability Density Functions (PDF), and Machine Learning (ML) models. Our goal is to demonstrate the feasibility of using machine learning algorithms on personality trait data to predict humans' responses to open-ended commonsense questions. We assess the performance of the {\it PerSense} algorithms for personality assessment by conducting an experiment focused on Neuroticism, an important personality trait crucial in mental health analysis and suicide prevention by collecting data from a diverse population with different Neuroticism scores. Our analysis shows that the algorithms achieve comparable results to the ground truth data. Specifically, the PDF approach achieves $97$\% accuracy when the confidence factor, the logarithmic ratio of the first to the second guess probability, is greater than $3$. Additionally, ML approach obtains its highest accuracy, $82.2$\%, with a multilayer Perceptron classifier. To assess the feasibility of commonsense reasoning analysis, we train ML algorithms to predict responses to commonsense questions. Our analysis of data collected with $300$ participants demonstrate that {\it PerSense} predicts answers to commonsense questions with $82.3$\% accuracy using a Random Forest classifier. 
\end{abstract}
	
	}	

\maketitle
\IEEEdisplaynontitleabstractindextext

\IEEEpeerreviewmaketitle

	\section{Introduction}
	
	The demand for personalized services and adaptive technologies such as driver-less cars are projected to grow rapidly and will soon be integrated into our daily life. The communication of these technologies with human-beings needs to be proper and well-trained in order to serve, explain the facts, and understand different mindsets better. Automating the process of understanding individuals' priorities and their perception of the world can potentially be used to address specific needs of individuals and to establish efficient interactions among human and technologies. Human commonsense reasoning involves human factors and personality traits in the process of judgment and decision making, and it goes beyond the conventional logical deductive or experimental inductive reasoning. These human factors include perceptions, desires, herd mentality, mental stability, prejudice or other biases, behavior traits, and personal/ethical/religious values. Therefore, commonsense reasoning is difficult to formulate for smart technologies. As an example, the complexity of predicting stock market fluctuations is due to human factor or market psychology, and is not well understood by human-beings \cite{heike}. Similarly, the rise and fall of political figures in a democratic society is not necessarily merit-based and not even deterministic and time-invariant \cite{krueger}. In marketing, other than functionality and appearance of a product, consumers' evolving taste contributes to a brand popularity and its success or failure. Understanding the limitations and ethical consequences of commonsense reasoning is paramount and requires development of machine commonsense reasoning.
	
	In philosophy, the distinction between something reasonable and not reasonable is highly challenging to formulate, especially when a person should choose between two morally questionable acts \cite{foot1967problem}. The trolley problem, which is a well-known ethical thought experiment discussed by Philippa Foot, is an example of a highly challenging task. In the trolley problem, it is presumed justifiable for a driver of a runaway trolley that is moving towards five railroad workers to divert the trolley to a branch where only one worker is on the track. 86\% of participants in the survey believed saving five lives is more important than saving a single life \cite{thomson1985trolley}. However, a surgeon may not kill a patient or even let the patient die to distribute the patient's vital organs to five other patients that would otherwise die. The math seems identical, sacrificing one to save five, but one decision seems reasonable and justifiable to many people and the second case is not. The hypothetical trolley problem can be extended to realistic scenarios with driver-less cars \cite{goodall2016can}. A driver-less car should be equipped with commonsense reasoning for simply maximizing the gain and minimizing the loss in its decision making process.
	
	The complexity of commonsense reasoning problem can be further explained by IBM Watson, the question answering computing system that defeated Jeopardy champions in 2011. Watson had special tools for analyzing natural language and had access to 200 million pages of information including the Wikipedia encyclopedia with no Internet access. Jeopardy questions were facts not human perceptions. However, in response to the question that in which US city, the largest airport was named for a World War \RNum{2} hero and its second largest airport was named for a World War \RNum{2} battle? Watson answered: ``What is Toronto?''.  The answer was a very uneducated guess as Toronto is not even located in US! The supercomputer had no human-like commonsense when answered this question. Obviously, not all different scenarios can be programmed or thought through ahead of time.  If we ask the same question from a large group of children, the likelihood of hearing ``Toronto'' as the answer is projected to be extremely low, even though they probably will not know the answer. Although IBM Watson easily defeated the Jeopardy champions, the real problem arises when IBM Watson is asked about human perceptions regarding issues that are not necessarily factual or for which there exists multiple correct answers for one question. As an example, it is trivial to list the name of all animals which were mentioned in the Bible. However, it is challenging for Watson to guess the most common answer of general public in US to this question. 
	
	In this paper, we hypothesize that ``commonsense reasoning'' is mostly grounded in human personality traits. To date, human personality traits have remained a virtually unexplored sources of information for understanding human beings' priorities and their perception of the world. There are numerous standardized questionnaires that have been developed for personality trait analysis \cite{roberts2006patterns,deneve1998happy,goldberg1993structure}. However, this traditional approach is based on the active participation of subjects in the conducted study. The validity of these approaches also depends on the subjects' honestly and dedication to correctly answer to the questions. Recently, a passive form of personality profiling is gaining interest in which the subjects' natural behavior is examined, rather than the choices that they intentionally make in filling questionnaires. The passive approach is projected to more accurately reflect the truth if the subjects' behavior over an extended period of time is available. The passive approach can also exploit vast information that is readily available on social media, from product reviews on Amazon to Twitter and Facebook posts \cite{correa2010interacts,hughes2012tale}. The passive form of personality profiling is projected to shed more light on human mindset by exploiting the massive information that is readily available on the Internet.

		\begin{figure*}[!t]
			\begin{center}
				\includegraphics[width=0.6\textwidth,keepaspectratio]{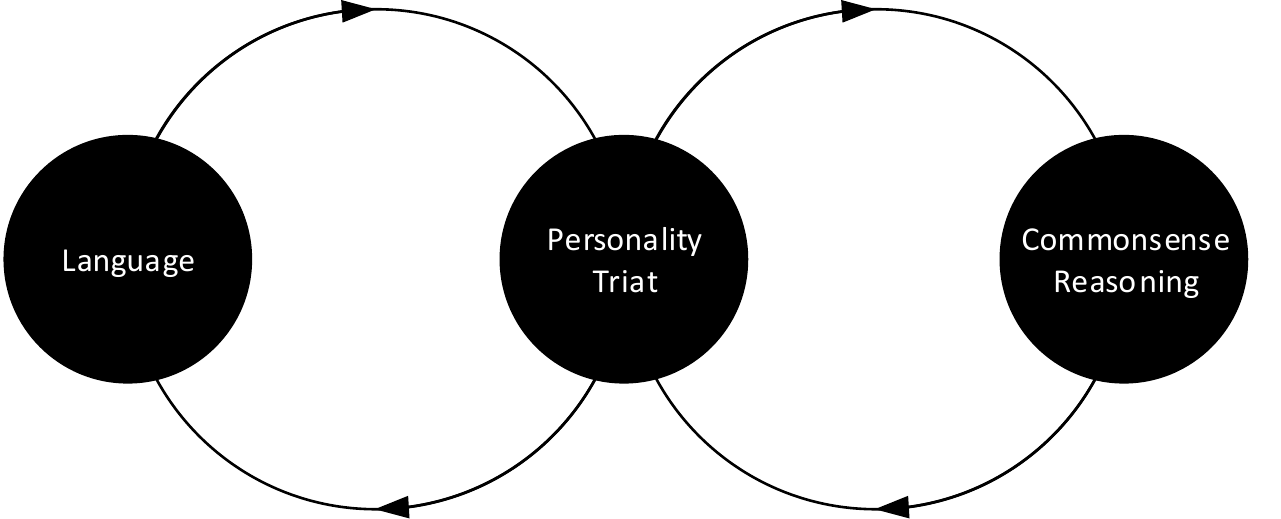}
				\caption{Utilizing personality traits for commonsense reasoning from language.}
				\label{figfiveFactor}
			\end{center}
		\end{figure*}
		
	In an entirely new direction, we propose {\it PerSense}, a novel framework for personality assessment and commonsense reasoning. The overall approach to depict the described hypothesis is shown in \figref{figfiveFactor}. The relationship between the personality and language is presumed to be present in two different directions and personality trait is also hypothesized to be related with commonsense reasoning. Therefore, our innovative work in this project aims to:
	
	\begin{enumerate}
		\item develop computationally-simple algorithms to estimate human personality traits based on user's digital footprint on social media,
		\item exploit the connection between personality trait and commonsense decision making, and
		\item demonstrate the feasibility of predicting human beings' response to open-ended questions that have no unique factual answer.
	\end{enumerate}
	
	Therefore, {\it PerSense} is a framework for commonsense reasoning based on personality traits. It first employs texts from different sources such as Amazon reviews for analyzing individuals' personality. We designed two low complexity approaches for estimating personality traits based on text data in {\it PerSense}. In the absence of a large scale survey based on standard personality traits questionnaires, the personality scores generated by the IBM Watson tool are used for ground truth labeling. These scores, which are real numbers between 0 and 1, are then utilized for statistical analysis and machine learning model training focused on personality analysis. Finally, we exploit human personality traits and their distributions in a society to develop machine commonsense reasoning. To this end, we use a machine learning based approach where we train the algorithms using data collected from hundreds of subjects through Amazon Mechanical Turk.

	\section{Related Work}
	
	In this section, we discuss related research on personality assessment via language and text. We also discuss previous research on commonsense knowledge and reasoning.
	
	\subsection{Personality Assessment}
	Personality has been found to significantly correlate with a number of real-world behaviors. It correlates with music taste: popular music tends to be significantly liked by extroverts, while people with a tendency to be less open to experience tend to prefer religious music and dislike rock music [32]. Personality also impacts the formation of social relations [36]: friends tend to be, to a very similar extent, open to experience and extrovert [35].
	
	Human personality traits, thoughts, emotions, and judgments are amongst the important topics in psychology \cite{buchheim2000relationship, gleser1959relationship}. Since the most universal way of expressing human feelings and thoughts is via language, it is reasonable to assume personality trait and language are related \cite{brown1968words,berry1997linguistic,mairesse2006words}. The lexical hypothesis, which is one of the well-established propositions in psychology, upholds this assumption \cite{goldberg1993structure}. The lexical hypothesis states that any language provides an unbiased source of different personality types. Thus, it postulates the  feasibility of extracting out personality types and psychometric features by examining the linguistic representation of people \cite{pennebaker2003psychological, antonioni1998relationship}. The measurement of psychometric features, which highlight individual differences \cite{pennebaker1999linguistic}, must be reliable in the sense that the measurement method and tools provide stable results across different circumstances. The measurement must also be valid in the sense that the results remain correct for various purposes \cite{beauducel2007impact}.
	
	The Big Five model or the five factor model (FFM) \cite{digman1990personality}, is a scheme of  personality traits classification \cite{de2002big, hendriks1997construction, deyoung2014openness}.  This model utilizes the evaluation of individuals' personality traits based on their ranking on five bipolar factors called OCEAN (Openness, Contentiousness, Extroversion, Agreeableness, and Neuroticism) \cite{mccrae1992introduction}, which is traditionally done using standard questionnaires. The validity of the Big Five model depends on the validity of the lexical hypothesis \cite{zeidner2000intelligence, deyoung2007between}. Accordingly, the validity of the lexical hypotheses, depends on its initial claim that personality is accurately represented within the structure of the language. There are interesting applications for the Big Five model. For example, It has been shown that extroverted individuals have more Facebook friends \cite{amichai2010social}. The users with high Neuroticism scores, on the contrary, post more on Facebook wall \cite{ross2009personality}. Another application of Big Five model is in prediction of job performance \cite{judge2013hierarchical}. 
	
	There are a few technical reports and papers on extracting out personality from lexicon which is called the Factor Analysis. There are different methods for factor analysis. The first and conventional methods where based on the personality traits description questionnaire \cite{drasgow1999innovations, cooper2010psychometric, shahar2004interactive}. It extracted out relevant information about personality from descriptive adjectives and different questions and phrases related to personality. The automatic recognition of personality traits was first presented in \cite{argamon2005lexical}. Recent developments in Artificial Intelligence (AI) explore new methods for automatic detection of linguistic features in text. For example, Roderick Hart developed the DICTION program in 2001, which counts the words in a text related to a particular theme \cite{hart2001redeveloping}. The DICTION was a word count program to reveal the verbal tone of political statements. The five master variables (activity, optimism, certainty, realism, and commonality) that provide a robust understanding of a text were used as a basic characterization of a text. A well-known tool for linguistic analysis is Linguistic Inquiry and Word Count (LIWC) \cite{chung2012linguistic, francis1993linguistic,pennebaker2003psychological}, which is an \textit{a priori} word-category approach. The LIWC method relies on an internal default dictionary (\textit{a priori} word-category) that defines which words should be counted in the target text files. For example, the word ``cried'' is part of five word categories: sadness, negative emotion, overall effect, verb, and past tense verb. 
	
	There are many available user text datasets such as social media, on-line reviews, and essays which can be utilized in inferring personality and building models for predicting an individual's behavior in society \cite{mcauley2016addressing,lambiotte2014tracking,lazer2009life}. In \cite{pentland2010honest}, an approach was proposed for capturing physical proximity, location, and movement. Later, these data were utilized for modeling human behavior. Phone and email logs can provide some patterns about human communication. A study was provided by Onnela et al., in which the communication patterns of millions of mobile phone users were utilized for observing the coupling between the strength of interactions and the structure of the network \cite{onnela2007structure}. In a work proposed by Eagle et al., physical proximities were collected over time for inferring cognitive relationships, such as friendship \cite{eagle2009inferring}. 
	
	The Internet is the main source for understanding the behaviors and connections of the people \cite{watts2007twenty,teevan2008people}. Personality has impacts on people's online interaction. In a research study, it is claimed that user's personality can be predicted using Facebook data, accurately \cite{golbeck2011predicting}. They utilized linear regression for predicting user's personalities. The input features included word counts, long words, words about time, health, etc. Mean absolute error for this prediction was 11\%. In another study, the impact of users' personality on Twitter activities was investigated \cite{quercia2011our}. In this research, the relationship between big five personality traits and different types of Twitter users was studied. Moreover, user's personality traits were predicted based on the number of followings, followers, and times the user has been listed on other's reading list. For predicting each personality trait, they utilized a 10-fold cross validation with 10 iterations using M5$\prime$ rules. The maximum root-mean-square error (RMSE) is $0.88$. In this work they utilized basic network properties in user' Twitter account. In our research, we collected data from different sources and text analysis was performed. Moreover, we utilized two different approaches, statistical and probabilistic, for predicting personality traits. In another research, they proposed that Instagram picture features can be correlated with personality \cite{ferwerda2015predicting}. They tried to predict personalities based on the way the pictures were taken and filters applied to them. Features were extracted from 22,398 pictures and mean values were calculated for each feature to create a measurement of central tendency. Later, a correlation matrix was created based on the central tendencies. Their findings suggested a relationship between users' personality traits and their posted pictures.
	
	\subsection{Commonsense Reasoning}
	Knowledge representation and reasoning are important problems in artificial intelligence. There are different rules and methods for knowledge representation \cite{russell2016artificial}. In order for a computer to act intelligently, it should have a wide knowledge representation of the world. Designing such a computer, requires information about the type of knowledge and the ways of obtaining it. The reasoning problem is a philosophical problem that can be solved clearly using intelligent systems. In reasoning programs, interaction with the world are through inputs and outputs that can be represented in different forms. One of these representations utilizes a set of sentences in the form of logical language. 
	
	Commonsense knowledge is the most general kind of knowledge that belongs to all people \cite{liu2004conceptnet}.
	There exist several properties such as defeasibility and context sensitivity \cite{liu2004commonsense}, which distinguish logic reasoning from human commonsense reasoning. Therefore, the use of natural language is considered to be the central component for commonsense knowledge representation \cite{sowa2000knowledge}. In order for a system to process the natural language similar to human beings, it must find the relations and information about the words and their meanings. In a study by Liu et al. a system called ConceptNet is proposed wherein natural language is explored as a tool for commonsense representation \cite{liu2004conceptnet}. In a work performed by George A. Miller, a framework called WordNet is proposed which benefits from the relationship between words \cite{miller1995wordnet}. In another work by Singh et al. a system called Open Mind is designed that obtains commonsense knowledge from people over the web \cite{singh2002open}. A calender with commonsense called SensiCal is presented by Mueller that generates warnings for avoiding errors \cite{mueller2000calendar}. For example, SensiCal sends a message to the user not to take a friend that is a vegetarian to a steakhouse. 
	
	\section{PerSense Framework Design}
	
\begin{figure}[!t]
			\begin{center}
				\includegraphics[width=0.45\textwidth,keepaspectratio]{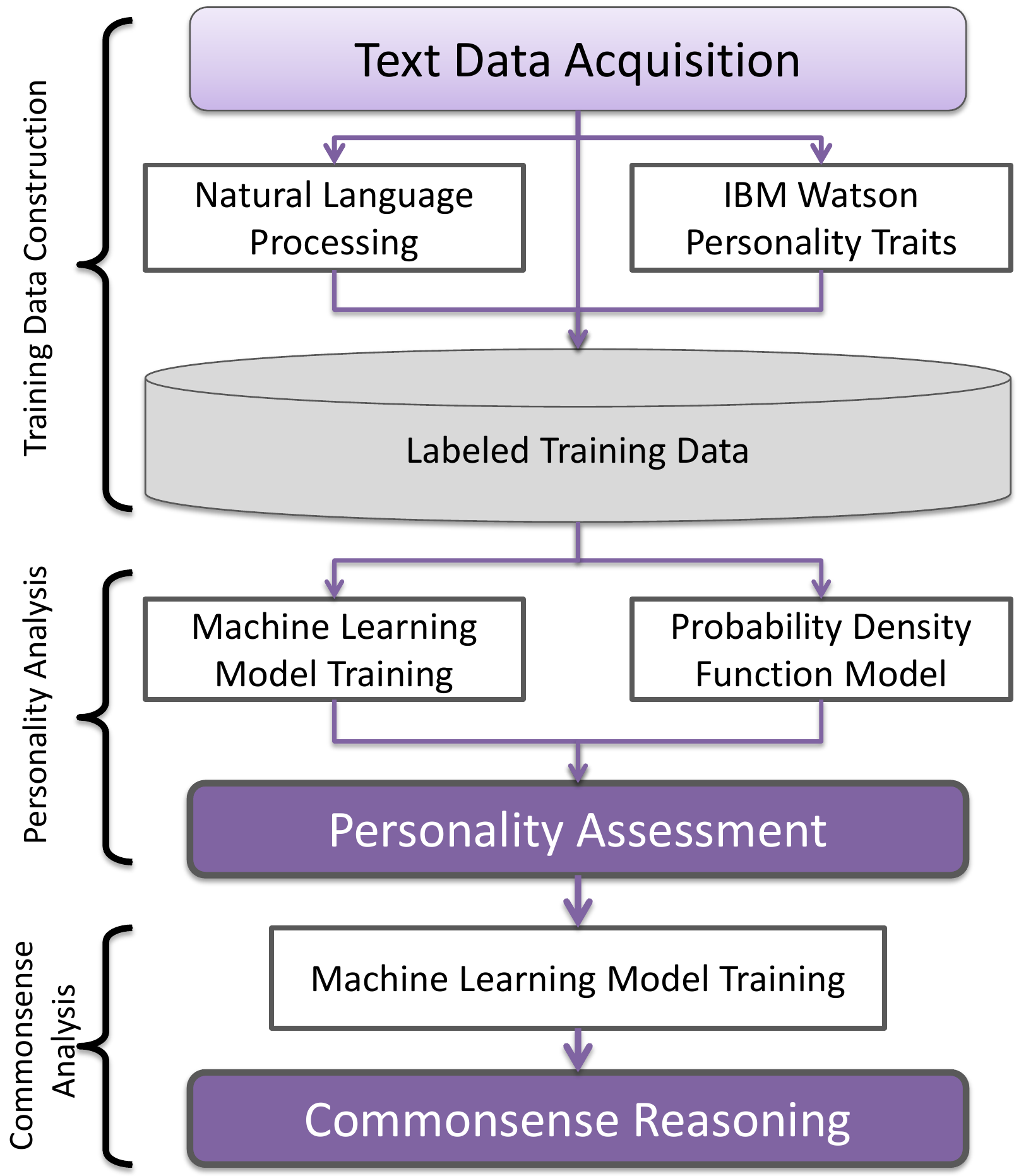}
				\caption{An overview of the {\it PerSense} framework.}
				\label{fig:system}
			\end{center}
		\end{figure}

	{\it PerSense} aims to perform commonsense reasoning based on an individual's personality. An overview of the {\it PerSense} framework for algorithm design is shown in \figref{fig:system}. Designing the {\it PerSense} framework takes three phases including 1) training data construction, 2) personality assessment, and 3) commonsense reasoning. In order to construct the training data, text data from different sources are gathered, natural language processing is applied on the data to extract important information from the text, and the personality traits associated with each text sample are scored using the IBM Watson Personality Insights engine. The text data, the information extracted using the NLP algorithms, and the scores achieved through the IBM Watson engine are then stored in a database. After gathering the labeled training data, {\it PerSense}  estimates personality traits for each text sample. Two models are proposed for estimating personality scores including 1) a probabilistic approach based on Aggregated Probability Density Functions (PDF); and 2) an ML-based approach that utilizes machine learning algorithms in a supervised fashion to learn a model that classifies text data into personality classes. After performing personality assessment, {\it PerSense} performs machine commonsense reasoning based on each individual's different personality traits. {\it PerSense} uses a supervised learning based methodology for commonsense reasoning. In what follows the computational algorithms utilized in 
	{\it PerSense} framework are explained in more detail.
	 
	\subsection{Probabilistic-Based Personality Assessment}
	Let us assume $N$ text samples have been processed and for the score $s$, a real number between 0 and 1 ( $0 \leq s \leq 1$), is associated with a given personality trait $T$, such as Neuroticism.  We assume $g(s_1,s_2)$ is the function that shows the number of text samples, among $N$ processed text samples, for which the personality trait score is between $s_1$ and $s_2$, where $s_1 \leq s_2$ ($g(0,1)=N$). The score interval $[0,1]$ is divided equally into $n$ sub-intervals of $[s_0,s_1]$, $[s_1, s_2]$, $\cdots$, $[s_{n-1}, s_n]$, where $s_0=0$ and $s_n=1$. In a uniform distribution of samples, $g(s_i, s_{i+1})= \frac{N}{n}$, where $i$ is any positive integer number between $0$ and $n-1$. In practice, it is unlikely to have a uniform distribution of samples as the chance of observing people with very high or very low scores is less than observing people with moderate scores. In any  statistical analysis, increasing the number of samples will increase the accuracy of the results. Consequently, the non-uniform distribution of samples will adversely impact the accuracy of the results. However, in a sufficiently large survey there will be a sufficient number of samples in any interval. 
	
	Any collected data in the $k$-th sub-interval, where $1 \leq k \leq n$, will be divided by the number of text samples in that interval ($g(\frac{k-1}{n}, \frac{k}{n})$) to generate a normalized metric. This metric indicates the likelihood of using a specific word (in this study an adjective) by authors whose personality trait score is in the $k$-th interval for the personality trait $T$. To achieve this objective, we create a histogram for each adjective with regard to a particular personality trait over personality trait intervals $[s_0,s_1]$, $[s_1, s_2]$, $\cdots$, $[s_{n-1}, s_n]$. For example, if a word $w$ occurred $x$ times in a text sample and the personality score for that text sample for a trait $T$ is $s_y$, which is in the $k$-th interval, i.e., $\frac{k-1}{n} \leq s_y \leq \frac{k}{n}$, a count of $x$ will be added to the number of times $w$ has been observed in the text samples authored by subjects whose scores are located in the $k$-th interval. 
	By processing a large number of text samples, i.e., $N, n \rightarrow \infty$ and normalizing the generated histogram by its integral from 0 to 1 to have an area equal to 1, the histogram will be converted to a probability density function, $f_{w,T}(s)$. The probability that an author that uses $w$ in her writing has a specific personality trait score between $\alpha$ and $\beta$ is calculated by $\int_\alpha^\beta f_{w,T}(s)ds$.
	
	It is equally important in probabilistic-based and ML-based approaches to find words that convey distinctive information regarding personality traits, i.e., their probability density function is not uniform, i.e., $f_{w,T}(s) \neq 1$, at least for one personality trait. The words with the smallest variance in their probability density functions are more selective and convey more information. By developing the database and creating the probability density functions for informative words, it becomes possible to analyze new test samples without relying on IBM Watson. Any new text sample is processed and its informative words are extracted. The related probability density functions for informative words, which have already been developed, are combined to create an aggregated probability density function (PDF) for each personality trait. The aggregated probability density functions are then used to estimate the personality trait of the author. 
	
	Let us assume, $w_1, w_2,  \cdots, w_m$ are $m$ informative words in a sample text, where $m$ is a positive integer and the number of distinct informative words can be less than $m$, which means some informative words can be used more than one time in a sample text. The corresponding probability density functions are $f_{w_1,T}(s), f_{w_2,T}(s), \cdots, f_{w_m,T}(s)$, respectively. The aggregated probability density function for the personality trait $T$ of the text sample, $\Phi_T(s)$, is calculated by the following equation.

	\begin{equation}
	\label{eq:2}
	\Phi_T(s)=\frac{\prod \limits_{i=1}^mf_{w_i,T}(s)}{\int_{x=0}^1\prod \limits_{i=1}^mf_{w_i,T}(x)dx}
	\end{equation}
	
	Here, we made a simplifying assumption that utilizing an informative word is a completely independent event and conveys no information regarding any other informative words in the text sample, not even regarding the same informative word if it has been used multiple times. In other words, the fact that the subject author used $w_1$ in the first sentence of the text sample, has no information about using $w_i$ in the text sample, where $i$ is an integer number between 1 and $m$. As an example, if  $f_{w_1,T}(s), f_{w_2,T}(s), \cdots, f_{w_m,T}(s)=1$ and convey no information, $\Phi_T(s)$ will also be equal to 1 and will convey no information regarding the personality trait $T$. The same technique can be expanded to additional informative words (verbs, pronouns, etc.) as well as additional text features, such as the use of passive and active voice, capitalization, punctuation, spelling or grammatical errors, sentence length, and word position.
	
	\subsection{Machine Learning for Personality Assessment}
\begin{figure*}[!t]
		\begin{center}
			\includegraphics[width=\textwidth,keepaspectratio]{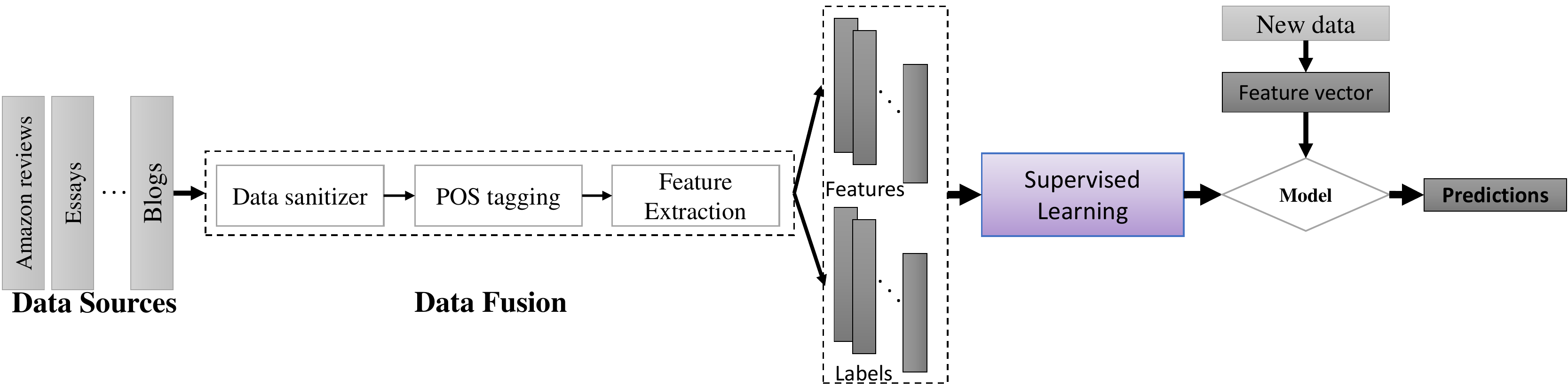}
			\caption{The pipeline of the personality assessment framework using supervised learning methodology.}
			\label{fig:ML}
		\end{center}
	\end{figure*}

	Alternatively, we developed a second approach for personality assessment based on supervised learning methodologies. There are a large number of methods for classification and regression in supervised learning research. Examples include tree-based, naive-based, support vector machines, etc. The classifiers and regressors do the predictions based on the input features to the system and their specific patterns. Supervised learning methodologies are useful techniques when a straightforward mathematical model cannot be extracted or building a statistical model is expensive. These techniques infer decisions based on a set of training data. The training data are divided into two sets: 
	
	\begin{enumerate}
		\item $X$ is a finite set of input features, 
		\item $Y$ is a finite set of output labels associated with each input feature instance, and 
		\item $F(X)$ is a function that maps input features to output labels, i.e., $X \overset{F(X)} \rightarrow Y$ 
	\end{enumerate}
	In this work, the input features are a set of adjectives utilized across the text samples and their frequencies in each text sample. The output labels are the personality scores associated with each text sample. $F(x_{i})$ is a classification algorithm that generates an output label $y_{i}$ as the predicted score/label for $sample_{i}$. The purpose of training is to find the mapping function from the input features to the output labels; therefore, if we feed a set of input features into our decision making function, it will provide a prediction of the output label. The schematic of the machine learning method is shown in \figref{fig:ML}.
	
	\subsection{Machine Learning for Commonsense Reasoning}
\begin{figure*}[!t]
		\begin{center}
			\includegraphics[height=45mm,keepaspectratio]{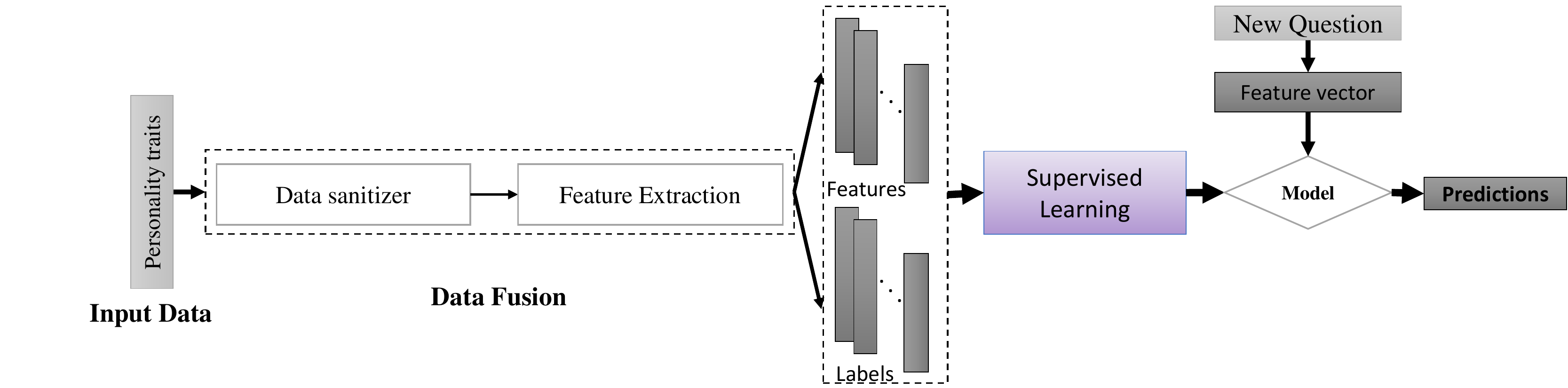}
			\caption{The pipeline of the commonsense reasoning framework.}
			\label{fig:CSR}
		\end{center}
	\end{figure*}

	We utilized a supervised learning based methodology for commonsense reasoning. Several classification and regression algorithms are utilized for modeling the problem. As mentioned previously, a supervised learning method derives the decisions based on the inputs that are given to the model. These inputs are represented as features in the context of machine learning modeling. For the purpose of commonsense reasoning, we designed a way of human-computer interaction wherein the input features are responses to a standard FFM personality trait questionnaire \cite{buchanan2005implementing}. This online FFM questionnaire was developed from the International Personality Item Pool (IPIP). A revised version of this was utilized for this work which was created with acceptable reliability and factor uni-vocal scales. FFM model was proved to be acceptable for use in internet-mediated research via different experiments. We perform classification for each individual commonsense question. The predefined responses to commonsense questions are output labels for that question. The schematic of the supervised learning method utilized for commonsense reasoning is shown in \figref{fig:CSR}.
			
	\section{Experimental Setup}
	This section describes the steps that are taken for evaluating {\it PerSense}. In order to prevent propagation of errors from personality assessment algorithms to commonsense reasoning techniques, we evaluated these two algorithmic modules individually. The final dataset for personality analysis using ML contains $16,934$ data points, $345$ input features, and $8$ class labels for each personality trait. The dataset for commonsense reasoning consists of $300$ data points, $50$ input features, and depending on each commonsense question there are different numbers of input features ranging from 2 to 7. 
	
	In this work, for evaluating the proposed framework, two sets of experiments where designed. First experiment is designed for evaluating personality assessment. For this purpose IBM Watson's text-based personality trait estimation tool was utilized for ground truth labeling. The accuracy of the proposed method was thus limited by IBM Watson’s accuracy. However, recruiting tens of thousands of volunteers to provide text samples and take standard personality tests was not a feasible option for this preliminary work. In the second part, standard personality tests results were utilized for commonsense reasoning, and the ground truth is generated by questionnaires filled through Amazon Mechanical Turk.  

    In what follows, the steps for data collection and experimental setup are explained in detail. We will discuss the obtained results based on these experiments in the next section.
	
	\subsection{Experiments for Personality Assessment}

	For evaluating the performance of the personality assessment algorithms, we construct a dataset of text samples to evaluate both approaches (i.e., PDF and ML) by utilizing existing texts from Amazon movie reviews \cite{mcauley2016addressing}, stream-of-consciousness essays \cite{humphrey1958stream}, and blog posts from sites hosted by WordPress. The mined text sample dataset contained $17,128$ data points. These text samples are given a score for each personality by the IBM Watson Personality Insights, a commercialized text-based personality assessment tool running on a big supercomputer and providing other fee-based services. In this work we utilized the IBM Watson service for assessing the personality characteristics based on a user's digital footprint and communications \cite{verma2009server}. This tool has limitations on the text inputs. Moreover, some of the adjectives are not descriptive due to having a low frequency in the samples. Therefore, the text samples that don't meet the following criteria are filtered out: 
	
	\begin{enumerate}
		\item text samples should be in English;
		\item the number of words in each sample should be greater than 600; and
		\item each adjective frequency should be greater than 25 throughout all the samples.
	\end{enumerate}
	The mined text sample dataset contained $16,834$ data points after data filtering.
	
	The scores given by IBM Watson are real values in the range of $0$ to $1$, with $0$ being the lowest in that personality trait. These scores are used as ground-truth labeling for testing our PDF and ML approaches for text-based personality trait assessment. The collected data were stored in a local database that consists of two tables. The first table has the list of all the adjectives, their frequencies, and a vector of scores associated with the samples that they were found in. The second table has all the text samples, a vector of adjectives and their frequencies in that sample, and the scores given by IBM Watson. Input features for training models for personality assessment are frequencies associated with each adjective. The output labels are the scores given to each text sample by IBM Watson.
	
	To build the training data in {\it PerSense}, the aggregated data are passed through Natural Language Processing (NLP) tools for extracting important data and their frequencies. For this purpose, the Natural Language Toolkit (NLTK) \cite{bird2004nltk} is utilized, which uses large bodies of linguistic data, or corpus. The NLTK is written in Python and distributed under the GPL open source license. We utilized existing corpora in the NLTK which are large text bodies for aggregating a list of existing adjectives, such as Gutenberg, web and chat text, Brown, and Reuters. A total of $26,414$ adjectives were found in the corpora. Utilizing this tool, we tagged the sentences in all of the text samples and extracted the adjectives and their associated frequencies in the text samples. For the purpose of ground-truth labeling, the dataset is analyzed by IBM Watson for assigning personality scores to each text sample. The data obtained from these two sources, NLP and IBM Watson, are then stored in a database. In what follows, the dataset for each personality assessment approach and statistics of the data are explained in more detail.
	
	\subsubsection{Distribution of Scores} 
	To obtain ground-truth labeling, the dataset was fed to the IBM Watson and scores were assigned to each text sample in the dataset. Out of different personality traits, OCEAN traits were extracted from the results. The resulted scores by IBM Watson were divided into 10 class labels in order to see the distribution of data. As shown in Table 2, Openness score for 90\% of the data was above 0.9, while the remaining 10\% of the data had a score between 0 and 0.9. Although all personality traits are equally important, the most uniform distribution of the mined samples was observed for Neuroticism. This personality trait is important in mental health care and suicide prevention. Therefore, our text-based analysis was focused on this personality trait.
	
	\begin{table}
		\centering
		\small
		\caption{The distribution of the data for each class label.}
		\begin{tabular}[ht]{|c||c|c|c|c|c|} \hline
			\textbf{Labels} & \textbf{O} & \textbf{C} & \textbf{E} & \textbf{A} & \textbf{N} 
			\\\hline
			\newline
			\textbf{0-0.1} & 0.3\% & 35\% & 31\% & 60\% & 6\%
			\newline
			\\\hline
			\textbf{0.1-0.2} & 0.6\% & 26\% & 20\% & 17\% & 6\%
			\newline
			\\ \hline
			\textbf{0.2-0.3} & 0.8\% & 15\% & 15\% & 8\% & 11\%
			\newline
			\\ \hline	
			\textbf{0.3-0.4} & 1\% & 9\% & 11\% & 4\% & 17\%
			\newline
			\\ \hline											
			\textbf{0.4-0.5} & 1\% & 5\% & 8\% & 3\% & 18\%
			\newline
			\\ \hline									
			\textbf{0.5-0.6} & 1\% & 4\% & 5\% & 2\% & 16\%
			\newline
			\\ \hline									
			\textbf{0.6-0.7} & 1\% & 2\% & 4\% & 2\% & 11\%
			\newline
			\\ \hline									
			\textbf{0.7-0.8} & 2\% & 2\% & 3\% & 1.5\% & 7\%
			\newline
			\\ \hline									
			\textbf{0.8-0.9} & 2\% & 1\% & 2\% & 1\% & 4\%
			\newline
			\\ \hline									
			\textbf{0.9-1} & 90\% & 1\% & 1\% & 0.4\% & 3\%
			\newline
			\\ \hline																					
		\end{tabular}	
		\label{tbl:distribution}
	\end{table}

	\subsubsection{Dataset Preparation for PDF Analysis} 
	
	While $26,414$ adjectives were obtained from the NLTK corpora, only $17,771$ were used in our text samples. At first, we divided the personality scores into 10 subsets ($n=10$). The IBM Watson scores that were below $0.1$ and higher than $0.9$ seemed dubious and were typically from short text samples. Therefore, we limit the analysis to scores between $0.1$ (10\%) and $0.9$ (90\%). The labels for each subset are shown in ~\tblref{pdfscores}. 
	
\begin{figure}
		\captionsetup[subfigure]{justification=centering}
		\centering
		\begin{subfigure}{0.4\textwidth}
			\includegraphics[width=\textwidth]{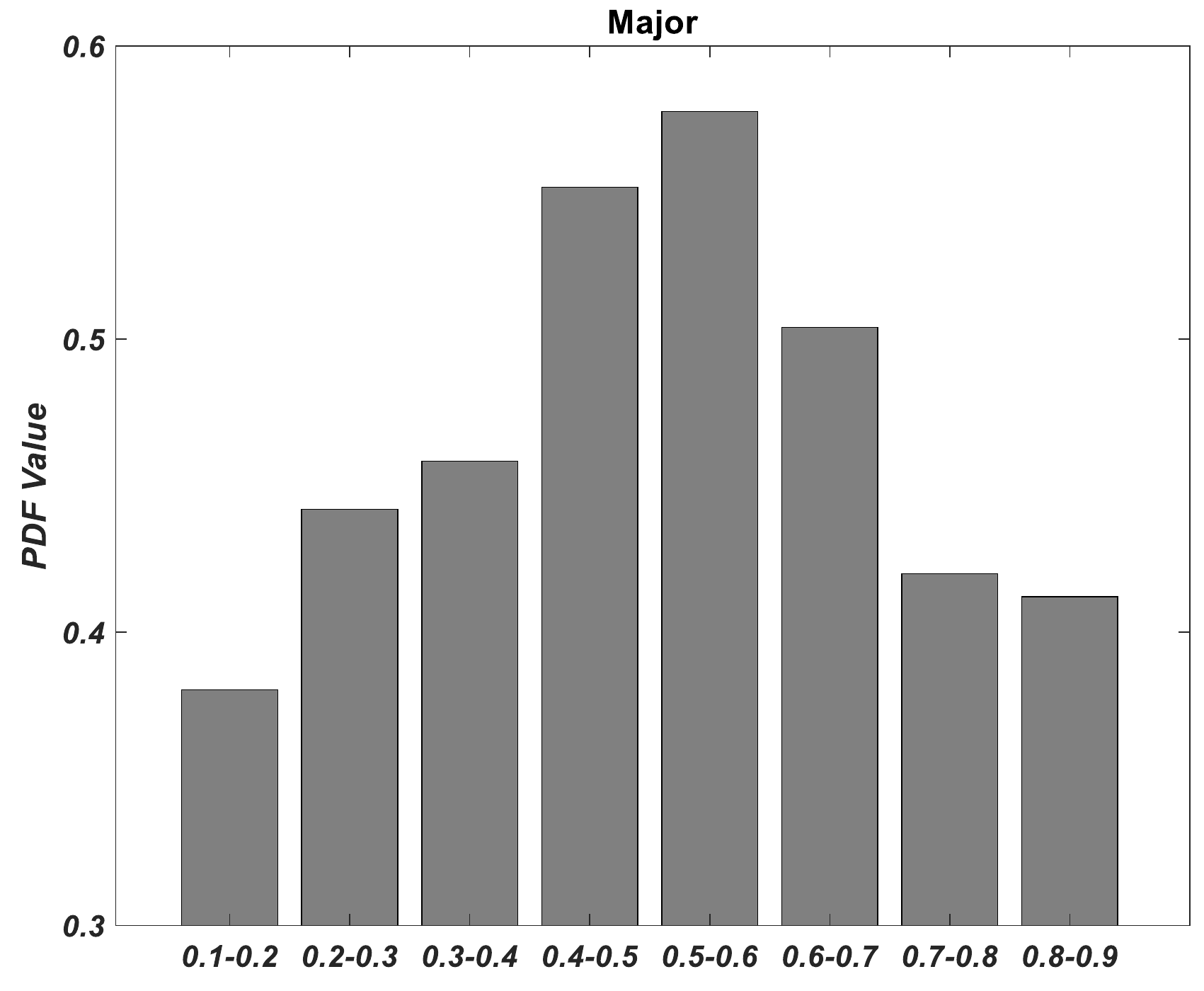}
			\caption{}
		\end{subfigure}
		\begin{subfigure}{0.4\textwidth}
			\includegraphics[width=\textwidth]{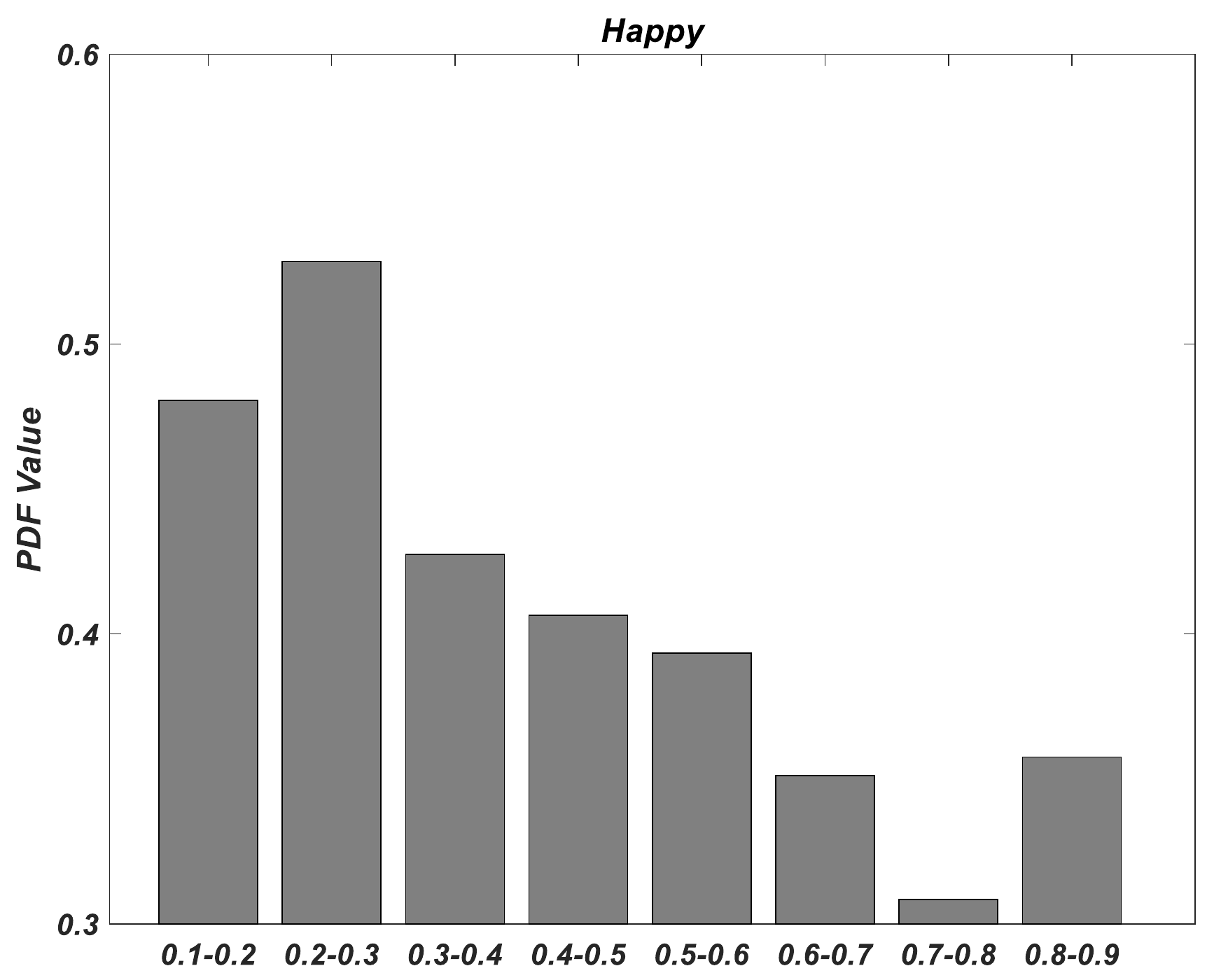}
			\caption{}
		\end{subfigure}
		\caption{The PDF values for two adjectives and the corresponding Neuroticism score: (a) Major --- (0.4-0.5), and (b) Happy --- (0.2-0.3).}
		\label{fig:adj1}
	\end{figure}

	The frequency of all the adjectives were computed separately for each sub-interval. The adjectives with frequencies less than $300$ over all the text samples were counted as unreliable data as there were zeros in the sub-intervals, resulting in $\Phi(s)=0$ in the related sub-interval based on (\ref{eq:2}). The final number of adjectives used in this study was $868$. \figref{fig:adj1} shows examples of PDF vectors for two adjectives: {\it Major} and {\it Happy}.
	
	After forming probability density function vectors for each adjective, we calculated the aggregated probability density function (PDF) for each text sample. The text samples with less than 1000 and more than 6000 words were eliminated. Text samples with less than 1000 words were too short to have reliable results using IBM Watson. A few text samples larger than 6000 words, up to one million words, were mined. These samples turned out not to be a usual text sample written by a subject reviewing a product or commenting about something and thus were discarded.   
	
\begin{figure}[!t]
		\begin{center}
			\includegraphics[width=0.5\textwidth,keepaspectratio]{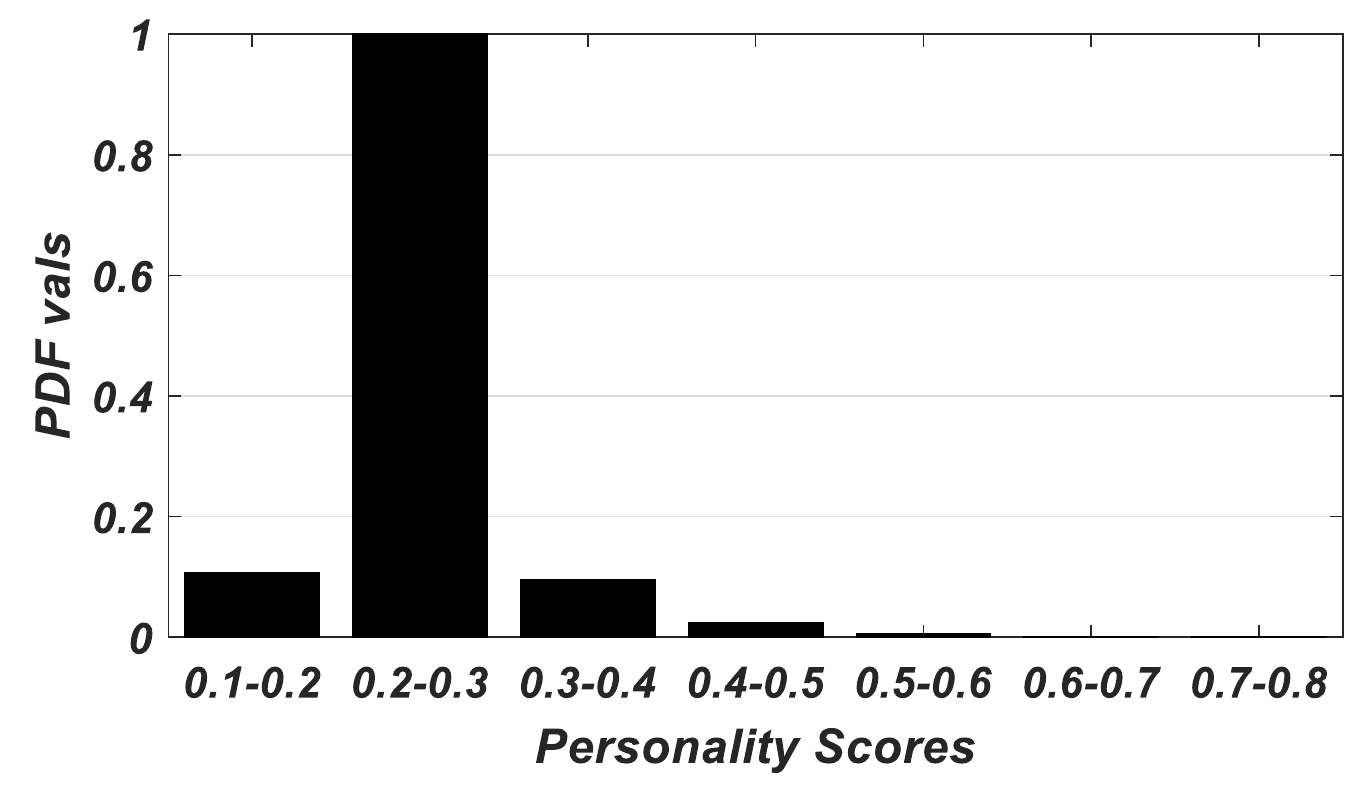}
			\caption{Neuroticism PDF for a specific text sample.}\label{fig:pdfEx}
		\end{center}
	\end{figure}

	Furthermore, we noted that we have more samples at the center and fewer samples at the corners, so the $g$ function was not uniform. In order to calculate each text sample's PDF vector, we extract the adjectives in each text sample and multiply their vectors. Finally, the PDF values are scaled to have an integral equal to one. Then the peak of each vector shows the personality score for that text sample. An example of a PDF output for Neuroticism is shown in \figref{fig:pdfEx}. This figure shows that the Neuroticism score for this specific text sample is in the range of $0.2$ to $0.3$.
	
	\begin{table}
		\centering
		\small
		\caption{The division of the personality scores and labels for each section.}
		\begin{tabular}[ht]{|c||c|c|c|c|c|} \hline
			\textbf{Personality Score Range} & \textbf{Label} 
			\\\hline
			\newline
			0.1-0.2 & 0.15
			\newline
			\\ \hline
			0.2-0.3 & 0.25
			\newline
			\\ \hline	
			0.3-0.4 & 0.35
			\newline
			\\ \hline
			0.4-0.5 & 0.45
			\newline
			\\ \hline
			0.5-0.6 & 0.55
			\newline
			\\ \hline		
			0.6-0.7 & 0.65
			\newline
			\\ \hline		
			0.7-0.8 & 0.75
			\newline
			\\ \hline		
			0.8-0.9 & 0.85
			\newline
			\\ \hline					
		\end{tabular}	
		\label{tbl:pdfscores}
	\end{table}
	
	\subsubsection{Dataset Preparation for ML Analysis}
	The initial set of input features are the total number of adjectives that were identified through the NLTK. After tagging the adjectives throughout all the text samples, the number of adjectives came down to $3,834$. Thereafter, adjectives that had a frequency of less than 10\% of the total number of occurrences of all the adjectives throughout the data points were removed from the feature set. The number of adjectives after data filtering was $345$. Other than filtering the feature set, the data points were filtered as well. The number of data points before data filtering was $17,127$. The frequencies of adjectives in each data point were enumerated, and data points with fewer than 5.5\% of the total number of features were removed from the dataset. 
	
	The final dataset contains $16,934$ data points and \textbf{$345$} input features. Each $x_{ij}$ in the feature set is the number of adjective $i$ in data point $j$.

	\subsubsection{Regression and Classification Models}
	Classification is a form of predictive modeling and it is focused on learning a mapping function from input features to output labels. The classification model is used for predicting a discrete class label output. The regression model, on the other hand, is used for predicting a continuous set of class labels.
	
	The main motivation of this analysis is to quantify the correlation between adjective usage in each text sample and the personality scores associated with that sample. The correlation of the adjectives and the personality scores is estimated using regression models. As mentioned above, the dataset contains 345 features after the process of feature extraction and feature selection. The regression models utilized in this work are Support Vector Regression (SVR) with radial basis function (rbf) kernel \cite{basak2007support}, Linear Regression \cite{kutner2004applied}, and Random Forest Regression \cite{liaw2002classification}. 
	
	Moreover, we tried to predict the output labels using classifiers as well. Six popular supervised learning classifiers were utilized for modeling the personality predictor, namely Support Vector Machine (rbf, degree 2 polynomial, and linear kernels) \cite{hearst1998support}, Decision Tree \cite{safavian1991survey}, K-nearest neighbors \cite{peterson2009k}, Perceptron \cite{freund1999large}, Multi-layer Perceptron \cite{gardner1998artificial}, and Random Forest classifier \cite{pal2005random}.
	
	\subsection{Experiments for Commonsense Reasoning}
	
	In this section, the setup of the experiments for commonsense reasoning based on the users' personality traits are elaborated. For the purpose of training a prediction model, data were collected using Amazon Mechanical Turk and Facebook. A test was designed for assuring the consistency of answers to survey questions. After passing the consistency test, we obtained 300 survey response and each response was considered as a data point. In what follows the data analysis steps are explained in detail.
	
	\subsubsection{Commonsense Dataset} For building the dataset for commonsense reasoning, we designed a survey with two parts. The first section of the survey carries $50$ personality trait questions. This questionnaire contains standardized five factor personality test. The responses are in the form of a Likert scale which is commonly involved in research studies that employs questionnaires or surveys: very inaccurate (1), moderately inaccurate (2), neither inaccurate nor accurate (3), moderately accurate (4), very accurate (5) \cite{allen2007likert}. The first set of questionnaire (personality questions) are input features for training commonsense reasoning models. In this article, input features are a set of scores given to each personality question by the user. There are $5^{50}$ possible outputs for standard personality questionnaire. Questions are expected to be independent and orthogonal, resembling eigenvectors in a vector space, each targeting a unique dimension of human personality. The FFM scores, historically were designed to be concise and handy for general statements not necessarily an adequate tool in big-data era and do not convey equal information for common sense analysis. 

    The second section of the survey contains $67$ questions, $75$\% from the game show Family Feud \cite{mcclintock1993family}, and $35$\% developed in our lab. The developed questions are based on Samsung's interest, popular political questions, and health-care problems. A model is trained for each commonsense question; therefore, the output labels are the answers to each question (e.g. Question: ``If money is not an issue, you will buy a driver-less car ...'', Answers: 1. Right away, 2. Only when 25\% of cars are driver-less 3. Only when you have no other option 4. Only when 50\% of cars are driver-less).
	
	The distribution of class labels was calculated for each question. If the class labels were not well distributed for a question, we tried to merge related labels and make the results uniformly distributed. The labels for each question were merged based on the relevance of each label to another on for a particular question. This relevance of each label to another one was determined by perception of the experimenter of those labels to be positive or negative. For instance in the question about free health-care (What is your opinion about creating united health care insurance by slightly increasing the sales/income tax?), two answers agreed and two answers disagreed with this. Therefore, the experimenter merged two negative answers and two positive answers.
	
	The label distributions of a subset of commonsense questions are shown in Table 3. This table shows examples of commonsense questions and the distribution of class labels for each question. As an example, row two in Table 3 (``President Trump's travel ban is''), shows a question with four possible choices (1. ``Great and improves the safety of legal and patriotic US citizens'', 2. ``Not good, because it is discriminatory'', 3.``Good, but should include more countries'', 4. ``Not good, since not all Muslim people in the banned countries should be punished because of the Sept. 11 attack''). The distribution of responses to this question for each choice was 25\%, 17\%, 10\%, and 48\%, for each of the four above-mentioned questions. Based on the distribution and relevance of these options, responses for choices 1 and 3 were combined and responses selecting choices 2 and 4 were combined. 
	
	\begin{table*}
		\centering
		\small
		\caption{The distributions of the answers to some of the commonsense questions.}
		\begin{tabular}[ht]{|m{10 cm}||c|c|} \hline
			\textbf{Question} & without fusion&with fusion
			\\\hline
			{\small Name a subject that sons discuss with their fathers rather than with their mothers.} & 5\%,33\%,46\%,16\%,&54\% ,46\%
			\\\hline
			{\small President Trump's travel ban is } & 25\%,17\%,10\%,48\%,&35\% ,65\%
			\\\hline
			{\small You share values with } & 26\%,39\%,19\%,16\%,&26\% 39\%,35\%
			\\\hline
			{\small What is your opinion about creating united healthcare by insurancing the sales/income tax? } & 54\%,20\%,14\%,12\%,&54\% 46\%
			\\\hline
		\end{tabular}	
		\label{tbl:distribution2}
	\end{table*}
	
	\subsubsection{Prediction Model} 
	Eight popular supervised learning classifiers were utilized for developing a machine commonsense reasoning model through human-computer interactions: three Support Vector Machine with three different kernels (rbf, degree 2 polynomial, and linear), Decision Tree, K-nearest neighbors, Perceptron, Multi-layer Perceptron, and a Random Forest classifier.

	\section{Results}
	
	We conducted two sets of experiments for testing our proposed methods. The results of two personality analysis approaches utilized in this research were compared with IBM Watson personality insights. IBM Watson itself utilized the personality score that was obtained from personality survey, as ground truth labeling. This comparison is performed due to the similarity between our framework and IBM Watson personality insight service. The Watson Personality Insights service extracts information from text data in user's text messages, tweets, and posts using linguistic analysis and performs personality analysis. In what follows, we demonstrate the detailed results for each part of the data analysis.
	
	\subsection{Performance of PDF Approach}
\begin{figure}
		\captionsetup[subfigure]{justification=centering}
		\centering
		\begin{subfigure}{0.4\textwidth}
			\includegraphics[width=\textwidth]{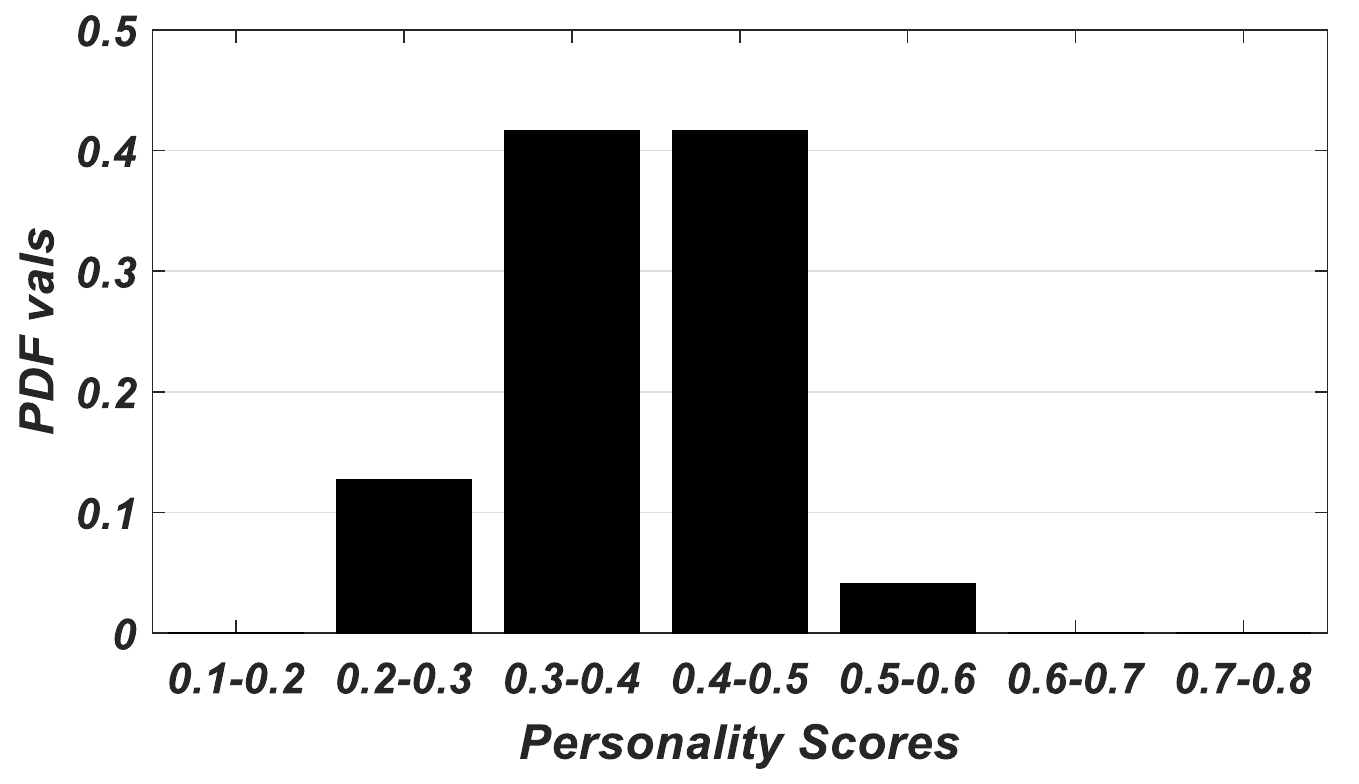}
			\caption{An example of a low confidence data point.}
			\label{fig:differentiable1}
		\end{subfigure}
		\begin{subfigure}{0.4\textwidth}
			\includegraphics[width=\textwidth]{HighC.pdf}
			\caption{An example of a high confidence data point.}
			\label{fig:differentiable2}
		\end{subfigure}
	\caption{Two data points with different confidence scores.}
	\end{figure}

	For evaluating the performance of the PDF approach, we calculated the PDF vector of each data point. Thereupon, the peak value of the vector was compared with the ground truth label.
	The results show that the Mean Absolute Error (MAE) and the Root Mean Square Error (RMSE) of the PDF method are 15.5\% and 19.5\%, respectively.
	The logarithmic ratio of the first to the second maximum was then calculated for each text sample for use as a measure of ``confidence''. The confidence scores were from 0 to 10 with 0 being the lowest confidence and 10 the highest.
	In some cases, the first and second peaks are comparable, while in other cases they are significantly different. An example of each is shown in \figref{fig:differentiable1} and \figref{fig:differentiable2}. The number of words in the text sample shown in \figref{fig:differentiable1} are $3,079$ and the score given by IBM Watson is $0.44$. One of the peaks given by PDF approach shows the Neuroticism value in the range of $0.4$ and $0.5$. In \figref{fig:differentiable2}, the number of words are $3,500$ and the Neuroticism score given by IBM Watson is $0.18$. The score given by the PDF approach is in the range of $0.2$ and $0.3$ which is close to IBM score.
	
\begin{figure}[!t]
		\begin{center}
			\includegraphics[width=0.5\textwidth,keepaspectratio]{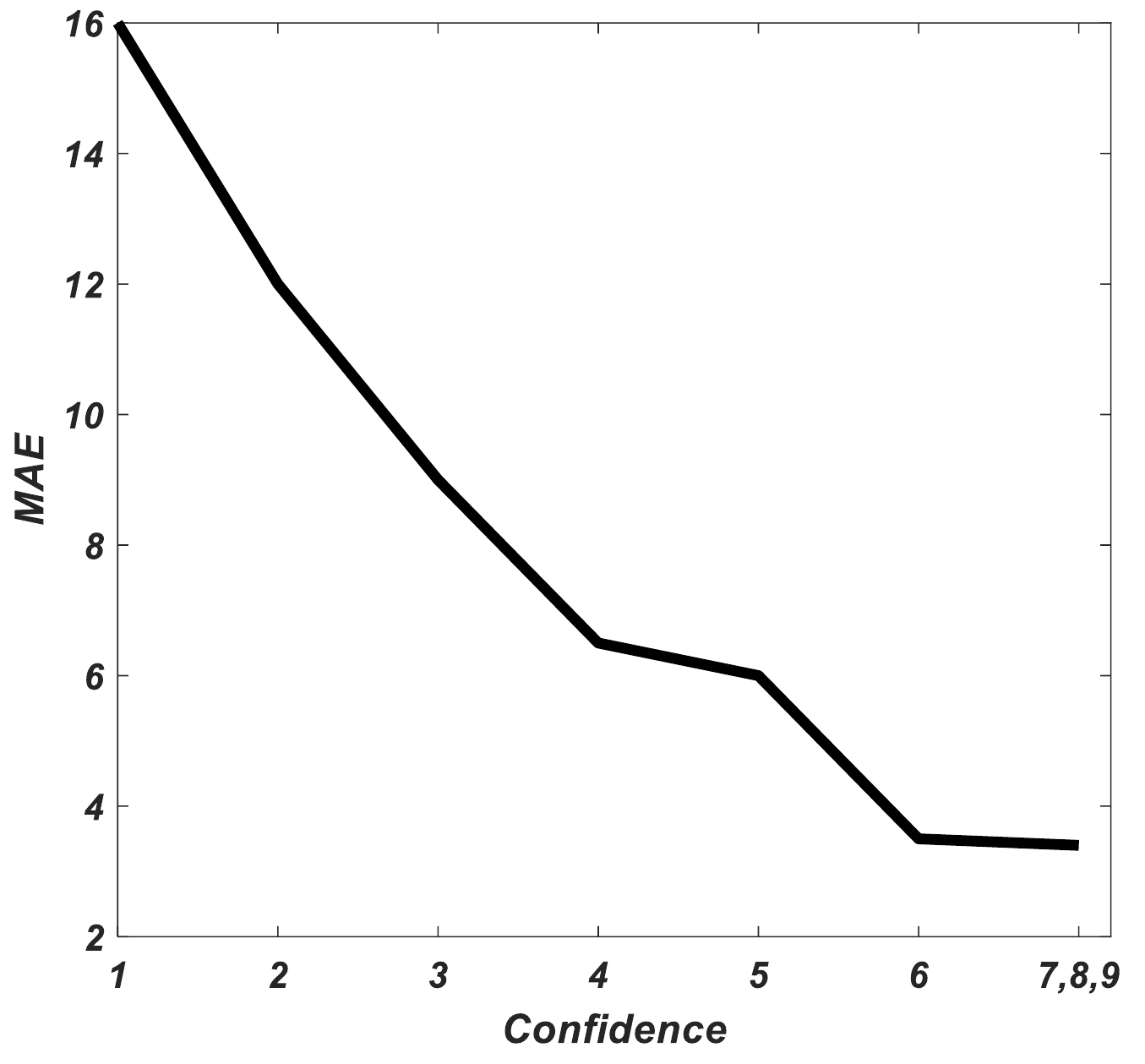}
		\end{center}\caption{The  Mean Absolute Error-Confidence plot shows by increasing the confidence metric the MAE drops below 4\%.}\label{fig:confidence}
	\end{figure}

	The MAE of the samples with confidence values more than two is 10.5\%. \figref{fig:confidence} shows the MAE based on different confidence levels.
	
	\subsection{Performance of ML-Based Personality Assessment}
	Since the amount of the mined data are large, we initially evaluated the accuracy using a train-test split (67\% training, 33\% testing). After choosing the best classifier, we used a 10-fold cross validation method for better accuracy evaluation. As it is shown in \tblref{distribution}, the data are not well distributed over all the scores. Therefore, we chose the most evenly distributed personality trait (Neuroticism) for performing our experiments and training models. For training ML classifiers, we utilized real value scores and converted them to integer percentages. While for training ML regressors we utilized real values as ground truth labeling, and output of the regressors were real values. Therefore, the accuracy of a classifier was measured based on the differences of the output labels which were numbers in range of 1 to 8, with the ground-truth labels generated by IBM Watson. The accuracy of a regressor was measured based on the difference between the actual score in range of 0-1 and the associated ground-truth labe.
	
	Scikit provides the root-mean-square deviation (RMSD) or root-mean-square error (RMSE) of each trained model which are frequently used as measures of the differences between actual values and the values predicted by a model. RMSD is based on the standard deviation of the predictions from ground-truth labels. While RMSE based on the average differences of predictions with ground-truth labels. The main purpose of this study is to predict a score as close as possible to the actual personality scores. Therefore, we defined an error margin for calculating the accuracy. If a predicted score is within the range of margin (-margin,+margin) away from the actual score, then it is counted as a correct prediction. Otherwise, it is counted as a mistake. The margin of error is set to 10\% for calculating the performance. It means if the score should be 90\% and the prediction is in the range of 80\% to 100\%, this is counted as a correct prediction. Results are shown in \tblref{performance}. The setting for each classifier and regressor is explained as follows:
	\begin{itemize}
		\item \textbf{SVM-Linear:} A support vector machine with linear kernel.
		\item \textbf{SVM-Polynomial:} A support vector machine with polynomial degree 2 kernel. 
		\item \textbf{SVM-rbf:} A support vector machine with Radial Basis Function kernel. This is one of the most popular kernels. For distance metric squared euclidean distance was used.
		\item \textbf{Decision Tree Classifier:} A supervised learning classifier that trains decision rules for classifying the data. In this work, the random state was set to zero.
		\item \textbf{KNN:} K-Nearest Neighbor is a supervised learning classifier. The classification is done based on the vote of its neighbors. The class label of each datapoint will be the most common class among its k nearest neighbor. In this paper, k was assigned to $5$.
		\item \textbf{Perceptron:} A supervised learning classifier, generally used for binary classification.
		\item \textbf{MLP:} Multi-layer Perceptron a artificial neural network classifier. We utilized a log-loss function as a solver. L2 regularization penalty is set to ${10}^{-5}$. The size of hidden layers was $15$.
		\item \textbf{SVR-rbf:} A support vector regressor with rbf kernel. Kernel coefficient was set to $0.1$ and the penalty parameter was set to ${10}^3$.
		\item \textbf{Linear Regression:} Attempted to model a relationship in data by fitting a linear equation to observed data.
		\item \textbf{RFR:} Random forest fits a number of classifying decision trees on the samples of the dataset and uses averaging to improve the predictive accuracy and control over-fitting. This is a random forest regressor with the maximum depth set to two.
		\item \textbf{RFC:} Random forest classifier with the number of trees set to 1000.
	\end{itemize}
	The highlighted cells are the lowest error values, obtained by the multi-layer perceptron.
	
	\begin{small}
		\begin{table}
			\centering
			\caption{The error metrics for each classifier}
			\begin{tabular}[ht]{|c||c|c|} \hline
				\footnotesize \textbf{Classifiers and Regressors} &\scriptsize Marginal Error & \scriptsize RMSE
				\\\hline
				\newline
				\textbf{\scriptsize SVM-Linear} &\scriptsize 37\% &\scriptsize 27\%
				\newline
				\\\hline
				\textbf{\scriptsize SVM-Polynomial} &\scriptsize 50\% &\scriptsize 47\% 
				\newline
				\\ \hline
				\textbf{\scriptsize SVM-rbf} &\scriptsize 39\% &\scriptsize 30\%
				\newline
				\\ \hline	
				\textbf{\scriptsize Decision Tree} &\scriptsize 50\% &\scriptsize 34\%
				\newline
				\\ \hline
				\textbf{\scriptsize KNN} &\scriptsize 45\% &\scriptsize 40\%
				\newline
				\\ \hline					
				\textbf{\scriptsize Perceptron} &\scriptsize 43\% &\scriptsize 29\%
				\newline
				\\ \hline
				\textbf{\scriptsize MLP} &\cellcolor{blue!25} \scriptsize 30\% & \cellcolor{blue!25} \scriptsize 22\% 
				\newline
				\\ \hline					
				\textbf{\scriptsize SVR-rbf} &\scriptsize 43\% &\scriptsize 26\%
				\newline
				\\ \hline		
				\textbf{\scriptsize Linear Regression} &\scriptsize 40\% &\scriptsize 24\%
				\newline
				\\ \hline	
				\textbf{\scriptsize RFR} &\scriptsize 46\% &\scriptsize 27\%
				\newline
				\\ \hline						
				\textbf{\scriptsize RFC} &\scriptsize 35\% &\scriptsize 27\% 
				\newline
				\\ \hline							
			\end{tabular}	
			\label{tbl:performance}
		\end{table}
	\end{small}
	
	\subsection{Performance of ML-Based Commonsense Reasoning} 
	
\begin{figure}[tbh!]
		\centering
		\begin{subfigure}[b]{0.4\textwidth}
			\centering
			\includegraphics[width=\textwidth]{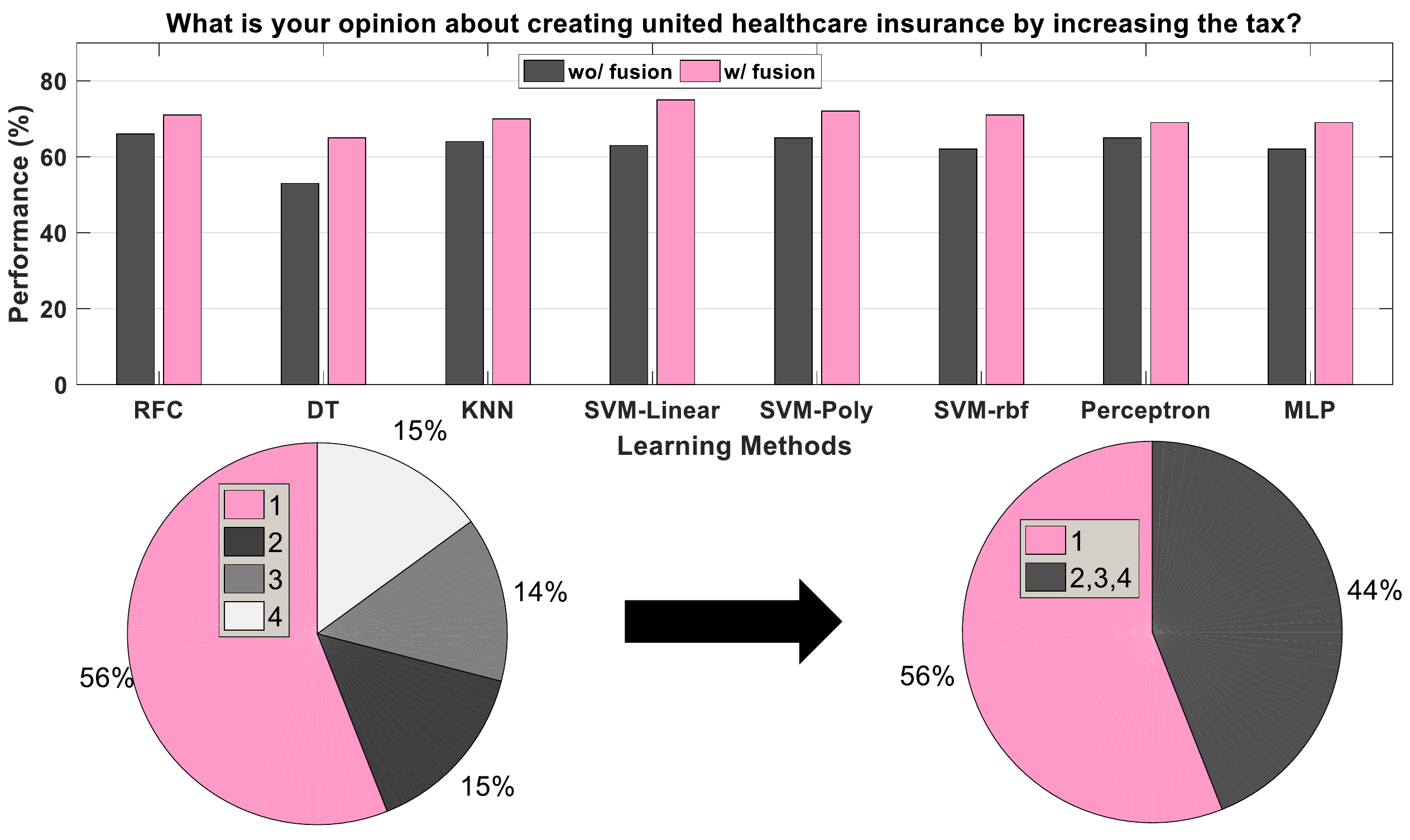}
			\caption{}
		\end{subfigure}%
		\qquad
		\begin{subfigure}[t]{0.4\textwidth}
			\centering
			\includegraphics[width=\textwidth]{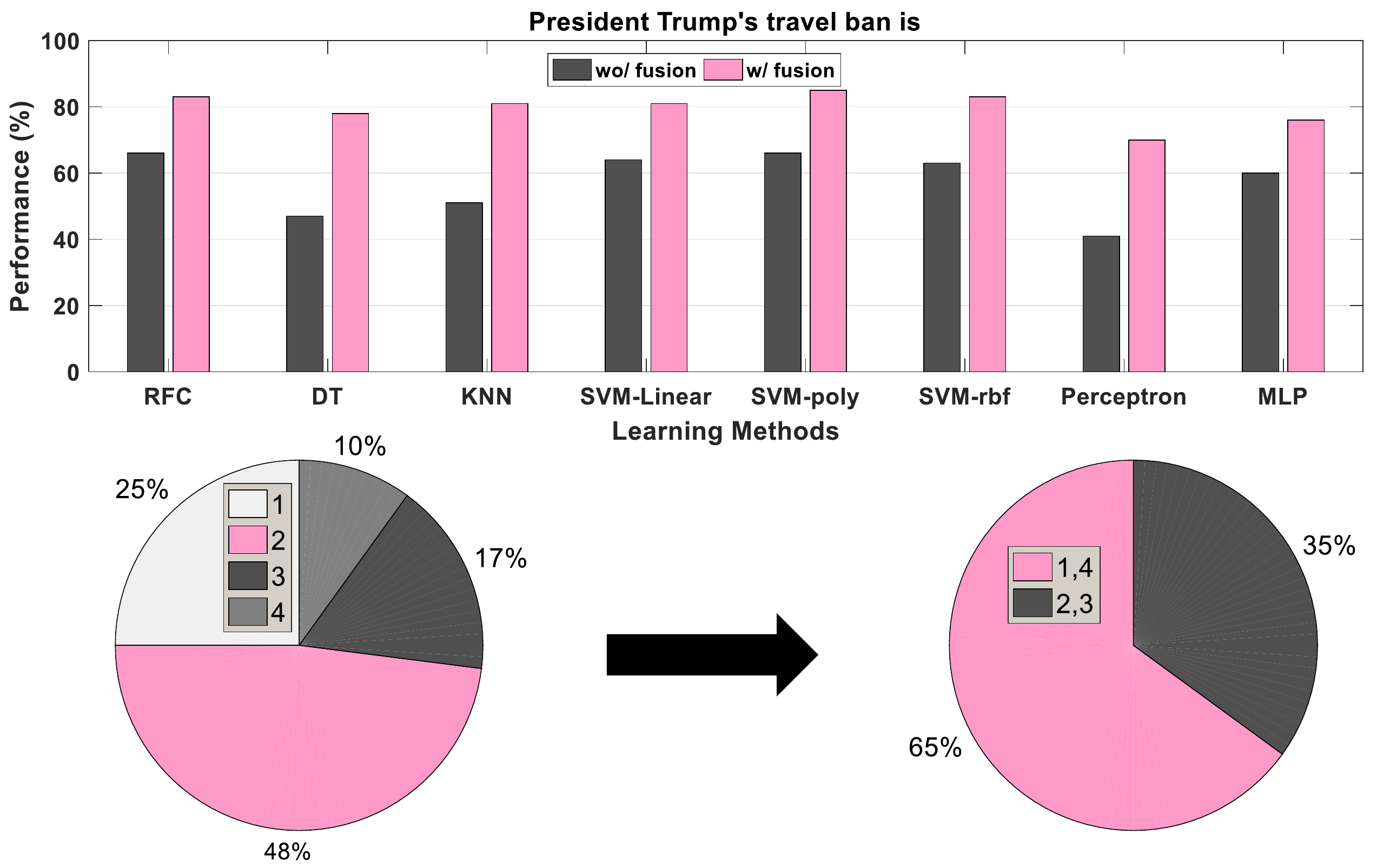}
			\caption{}
		\end{subfigure}
		\qquad 
		\begin{subfigure}[t]{0.4\textwidth}
			\centering
			\includegraphics[width=\textwidth]{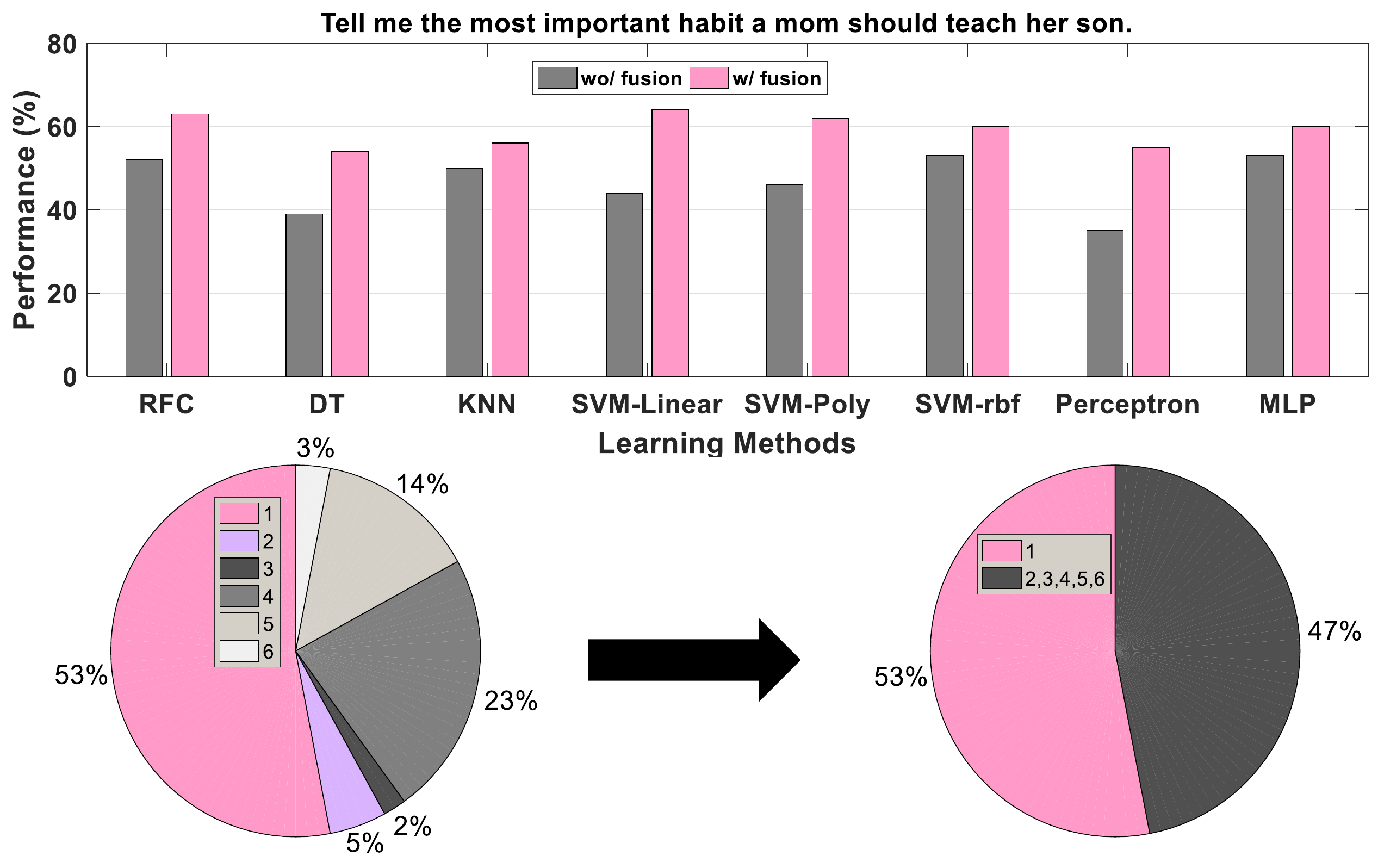}
			\caption{}
		\end{subfigure}
		\caption{Accuracy values of models before and after data fusion.}
		\label{fig:fusion}
	\end{figure}

	We initially calculated the correlation between each input feature and the output labels and the features with low correlations were filtered out in building the prediction models. This results in different feature sets for each commonsense question. A model was trained for each commonsense question using each training model ($8 \times 67$ models). For evaluating the performance, a 10-fold cross validation method was utilized. The non-uniform distribution of the class labels in some questions made the problem too easy and resulted in very high accuracies; therefore, these questions are not a good criterion for the evaluation of the models. On the other hand, some questions are well distributed between more than 3 class labels. Due to the low number of data points, these questions lead to significantly low accuracies.  The best criterion for evaluating the classifiers and regressors are the questions with uniform distribution among few numbers of class labels. We compared the accuracies of the classifiers for some of the questions before and after the data fusion. The results are shown in \figref{fig:fusion}. For instance, in \figref{fig:fusion}b the choices of the questions were: 1) Great and improves the safety of legal and patriotic US citizens (25\%); 2) Not good, since not all Muslim people in the banned countries should be punished because their fellow citizens committed the Sep.11 attack (17\%); 3) Good, but should include more countries (10\%);  and 4) Not good, because it is discriminatory (48\%). We merged options 1 with 3 and options 2 with 4, the distribution of the data became (65\% and 35\%). The average accuracy before and after the data fusion 52\% and 76\%, respectively.
	
			\begin{table}[H]
			\centering
			\caption{Confusion matrix for the commonsense question: ``What is your opinion about creating united healthcare insurance by increasing the tax?''}
			\begin{tabular}[ht]{|c||c|c|} \hline
			\backslashbox{\footnotesize{Correct}}{\footnotesize{Predicted}} & \textbf{Agree} & \textbf{Disagree} \\\hline
				\newline
				\textbf{Agree} & 60 & 20
				\newline
				\\\hline
				\textbf{Disagree} & 19 & 41 
				\newline
				\\\hline
			\end{tabular}	
			\label{tbl:healthscore}
		\end{table}
		\begin{table}[H]
			\centering
			\caption{Confusion matrix for the commonsense question: ``What do you think about Trump's travel ban?''}
			\begin{tabular}[ht]{|c||c|c|} \hline
				 \backslashbox{\footnotesize{Correct}}{\footnotesize{Predicted}} & \textbf{Agree} & \textbf{Disagree}
				\\\hline
				\newline
				\textbf{Agree} & 31 & 19
				\newline
				\\\hline
				\textbf{Disagree} & 9 & 81
				\newline
				\\\hline
			\end{tabular}	
			\label{tbl:trump}
		\end{table}
		\begin{table}[H]
			\centering
			\caption{Confusion matrix for the commonsense question: ``Tell me the most important habit a mom should teach her son?''}
			\begin{tabular}[ht]{|c||c|c|} \hline
				 \backslashbox{\footnotesize{Correct}}{\footnotesize{Predicted}} & \textbf{Respect} & \textbf{Others}
				\\\hline
				\newline
				\textbf{Respect} & 47 & 21
				\newline
				\\\hline
				\textbf{Others} & 30 & 41
				\newline
				\\\hline
			\end{tabular}	
			\label{tbl:son}
		\end{table}
	
	The confusion matrix for each commonsense reasoning question is shown in \tblref{healthscore} to \tblref{son}.

	\section{Concluding Remarks}
	The methodologies proposed in this article address important problems in the area of behavioral analysis through two novel approaches based on machine learning and probability density functions. The proposed system may have many different applications, including mental health care and suicide/homicide prevention. In this work, we applied our methodologies on personality assessment and commonsense reasoning.  Scores from IBM Watson Personality Insights were utilized for ground-truth labeling. However, our goal in this work was to match or surpass the performance of Watson with a system requiring far less computational resources. The proposed approaches achieved a comparable result to the ground truth labels. For the PDF approach with the confidence factor of higher than 3, the accuracy reached 97\%. The best results achieved by ML modeling was 88.2\% through the multilayer perceptron. In this article, we utilized 345 adjectives in order to train accurate models for personality analysis. 345 adjectives were outcome of our experiments based on the available text. We are planning on collecting more text data and include other words such as verbs, nouns, and pronouns in our experiments. Furthermore, by increasing the samples more adjectives can be added to the list. Although 345 input features are not ideal for training a model, in this work we targeted a low-complexity solution for running the proposed method on smartphones rather than an alternative for IBM Watson that needs a supercomputer to run.
	
	In this work, commonsense decisions were predicted based upon an individual's personality trait. For performing commonsense reasoning, we proposed an approach utilizing ML techniques. We collected data from $345$ people through Amazon Mechanical Turk and Facebook. Two sets of questions were provided. The first set contained the personality analysis questions which were utilized as input features for the commonsense reasoning problem. The second set contained the commonsense questions. Each commonsense question was modeled separately. In this context, commonsense questions had a number of reasonable and justifiable answers. Each answer was a class label for training a model. Based on the evaluations, Random Forest Classifier achieved 82.3\% accuracy in predicting the answers to some commonsense questions. We note that commonsense questions, which may target different political point-of-views, ethical priorities, marketing practices, and other personal choices, preferences, and habits, may have a number of reasonable and justifiable answers instead of a unique factual answer.
	
	In this study, we performed experiments to generate convincing evidences that the proposed low-complexity approach is in fact functional. However, more complex and comprehensive experiments are needed to assess large scale adoption of these technologies. We focused on a computationally simple method that can be easily implemented as a mobile application. Our goal in this study was not to develop an alternative approach to the IBM Watson tool, which uses a supercomputer to run. In the future, we plan to utilize a comprehensive set of word categories (e.g., verbs, nouns) throughout the text samples and use those data for design and development of our computational models. Moreover, in this study, we utilized one personality trait in our experiments. A future study will involve collecting data from various sources to incorporate all five personality traits in our experiments. 
	
	There were few sources of error in designing the algorithms. The number of subjects participated in the data collection was limited so that we could not cover a large number of personality traits at a large scale. To train commonsense reasoning models, we collected data through Amazon mechanical Turk; however, we were able to provide only a nominal amount of monetary incentive to the participants. This may have potentially impacted the quality of the collected data. Moreover, the number of subjects taking the survey was limited, and the survey results may have been affected by many factors such as personal biases. In our ongoing research, we are currently designing a data collection framework, which prompts the user with questions throughout the day, asking them to express their daily activities such as their experiences, their dietary intake, and interactions. The inputs to this system are in form of spoken language. The speech is then converted to text and utilized for personality analysis. For the commonsense reasoning analysis, we plan on recruiting more participants to train highly accurate models for commonsense reasoning.

\section{Discussion and Future Direction}

In this work  IBM Watson Personality Insight tool, which authors modeled by two low-complexity approaches, only provides 5 FFM scores instead of providing probable responses to a standard FFM questionnaire. However, there are $5^{50}$ distinct cases when a standard FFM questionnaire consisting of 50 questions and 5 possible answers per question is filled, thus the questionnaire conveys more information than only 5 FFM percentiles. Questions in a standard FFM questionnaire, are expected to be independent and orthogonal to each other meaning that each question targets a unique dimension of human personality and closely mimic Eigenvectors in an N-dimensional space. Authors built the commonsense reasoning scheme directly based on responses to the questionnaire. 

As stated, personalizing the algorithms and services in most efficient way has been the focus of many recent researches in human-centered applications \cite{active,ferwerda2015predicting,efficient,ma2019cyclepro}. Therefore, assessing the personality of humans would be a helpful and necessary step. This work is the first step towards integrating personality analysis and commonsense reasoning with all the limitations that research in academia has Nevertheless. However, it has achieved its objective to establish the connection between personality trait and common-sense reasoning. In this work, we have not achieved the ultimate goal to directly extract commonsense reasoning from text. To achieve this ultimate goal, one should exceed IBM Watson and need to have access to a really large number of text samples. Although, authors did not have access to the required resources; they highlighted the missing parts and will pursue these in their future direction.

Although this study clearly showed that commonsense reasoning is related to personality trait, the authors do/did not have the means to fill the FFM questionnaire directly from a text sample. Authors believe FFM scores, that are easier to use, convey less information than the questionnaire and therefore, IBM Watson as well as our method cannot rigorously revive the loss of information and fill the questionnaire using FFM percentiles. This is an interesting topic that can be explored in the future and requires a large number of text samples and responses to questionnaires to facilitate predicting responses to questions based on text features. 

Although applications need many words to learn personalities, it does not need to happen instantaneously. The users do not need to write an essay every time. The final product can use people’s voice, text, and other modalities and collect data overtime. As more users adopt the system and add more data to the applications, the performance improves. In our ongoing work, we are collecting data from users in real-time on daily basis by prompting them with specific questions throughout the day. For example, how was your day? What did you eat? Tell me about a movie you watched, a book you read, or a conversation you had. Then this data is utilized for personality assessment and commonsense reasoning. We expect that the amount of data collected over time will be an order of magnitude greater than what we have right now. Huge amounts of data will be available to improve the performance dramatically. 

In this work, the questionnaire is filled by a small number of users with different personality traits, rather than a large number of people in the society. The connection between a commonsense question and a specific feature in the personality trait is established using machine learning. In general, people in the society will not need to fill a questionnaire if their personality can be estimated by other means including their digital footprint. 

\bibliographystyle{IEEEtran}
\bibliography{refs}

	\newpage

\begin{appendices}
	
	\onecolumn
	\section{Survey Personality Questions}
	\begin{longtable}[b]{|c||m{1.5cm}|m{1.4cm}|m{1.5cm}|m{1cm}|m{1.5cm}|} \hline
		\footnotesize \textbf{I see myself as someone ...} &\scriptsize Very Inaccurate & \scriptsize Moderately Inaccurate & \scriptsize Neither Inaccurate nor Accurate  &\scriptsize Moderately Accurate  &\scriptsize Very Accurate
		\\\hline
		\footnotesize \textbf{who is generally trusting} &$\square$& $\square$ & $\square$  &$\square$  & $\square$
		\\\hline
		\footnotesize \textbf{Don't like to draw attention to myself} &$\square$& $\square$ & $\square$  &$\square$  & $\square$
		\\\hline
		\footnotesize \textbf{Believe in the importance of art} &$\square$& $\square$ & $\square$  &$\square$  & $\square$
		\\\hline
		\footnotesize \textbf{tends to find fault in others} &$\square$& $\square$ & $\square$  &$\square$  & $\square$
		\\\hline
		\footnotesize \textbf{pays attention to details} &$\square$& $\square$ & $\square$  &$\square$  & $\square$
		\\\hline
		\footnotesize \textbf{Shirks duties} &$\square$& $\square$ & $\square$  &$\square$  & $\square$
		\\\hline
		\footnotesize \textbf{who is generally trusting} &$\square$& $\square$ & $\square$  &$\square$  & $\square$
		\\\hline
		\footnotesize \textbf{who is reserved} &$\square$& $\square$ & $\square$  &$\square$  & $\square$
		\\\hline
		\footnotesize \textbf{has a good word for everyone} &$\square$& $\square$ & $\square$  &$\square$  & $\square$
		\\\hline
		\footnotesize \textbf{insults people} &$\square$& $\square$ & $\square$  &$\square$  & $\square$
		\\\hline
		\footnotesize \textbf{feels comfortable around people} &$\square$& $\square$ & $\square$  &$\square$  & $\square$
		\\\hline
		\footnotesize \textbf{dislikes self} &$\square$& $\square$ & $\square$  &$\square$  & $\square$
		\\\hline
		\footnotesize \textbf{has a little to say} &$\square$& $\square$ & $\square$  &$\square$  & $\square$
		\\\hline
		\footnotesize \textbf{has frequent mood swings} &$\square$& $\square$ & $\square$  &$\square$  & $\square$
		\\\hline
		\footnotesize \textbf{often feels blue} &$\square$& $\square$ & $\square$  &$\square$  & $\square$
		\\\hline
		\footnotesize \textbf{does not like art} &$\square$& $\square$ & $\square$  &$\square$  & $\square$
		\\\hline
		\footnotesize \textbf{cuts others to pieces} &$\square$& $\square$ & $\square$  &$\square$  & $\square$
		\\\hline
		\footnotesize \textbf{is the life of the party} &$\square$& $\square$ & $\square$  &$\square$  & $\square$
		\\\hline
		\footnotesize \textbf{believes that others have good intentions} &$\square$& $\square$ & $\square$  &$\square$  & $\square$
		\\\hline
		\footnotesize \textbf{avoids philosophical discussions} &$\square$& $\square$ & $\square$  &$\square$  & $\square$
		\\\hline
		\footnotesize \textbf{makes friends easily} &$\square$& $\square$ & $\square$  &$\square$  & $\square$
		\\\hline
		\footnotesize \textbf{gets chores done rightaway} &$\square$& $\square$ & $\square$  &$\square$  & $\square$
		\\\hline
		\footnotesize \textbf{accepts people as they are} &$\square$& $\square$ & $\square$  &$\square$  & $\square$
		\\\hline
		\footnotesize \textbf{relaxed, handles stress well} &$\square$& $\square$ & $\square$  &$\square$  & $\square$
		\\\hline
		\footnotesize \textbf{does a thorough job} &$\square$& $\square$ & $\square$  &$\square$  & $\square$
		\\\hline
		\footnotesize \textbf{does just enough work to get by} &$\square$& $\square$ & $\square$  &$\square$  & $\square$
		\\\hline
		\footnotesize \textbf{enjoys going to art museums} &$\square$& $\square$ & $\square$  &$\square$  & $\square$
		\\\hline
		\footnotesize \textbf{often down in the dumps} &$\square$& $\square$ & $\square$  &$\square$  & $\square$
		\\\hline
		\footnotesize \textbf{always prepared} &$\square$& $\square$ & $\square$  &$\square$  & $\square$
		\\\hline
		\footnotesize \textbf{has few artistic interests} &$\square$& $\square$ & $\square$  &$\square$  & $\square$
		\\\hline
		\footnotesize \textbf{not easily bothered by things} &$\square$& $\square$ & $\square$  &$\square$  & $\square$
		\\\hline
		\footnotesize \textbf{knows how to captivate people} &$\square$& $\square$ & $\square$  &$\square$  & $\square$
		\\\hline
		\footnotesize \textbf{is not interested in abstract ideas} &$\square$& $\square$ & $\square$  &$\square$  & $\square$
		\\\hline
		\footnotesize \textbf{makes plans and sticks to them} &$\square$& $\square$ & $\square$  &$\square$  & $\square$
		\\\hline
		\footnotesize \textbf{doesn't see things as they are} &$\square$& $\square$ & $\square$  &$\square$  & $\square$
		\\\hline
		\footnotesize \textbf{is outgoing and sociable} &$\square$& $\square$ & $\square$  &$\square$  & $\square$
		\\\hline
		\footnotesize \textbf{doesn't talk a lot} &$\square$& $\square$ & $\square$  &$\square$  & $\square$
		\\\hline
		\footnotesize \textbf{gets back at others} &$\square$& $\square$ & $\square$  &$\square$  & $\square$
		\\\hline
		\footnotesize \textbf{tends to vote for the liberal candidates} &$\square$& $\square$ & $\square$  &$\square$  & $\square$
		\\\hline
		\footnotesize \textbf{has an active imagination} &$\square$& $\square$ & $\square$  &$\square$  & $\square$
		\\\hline
		\footnotesize \textbf{seldoms feel blue} &$\square$& $\square$ & $\square$  &$\square$  & $\square$
		\\\hline
		\footnotesize \textbf{carries out the plans} &$\square$& $\square$ & $\square$  &$\square$  & $\square$
		\\\hline
		\footnotesize \textbf{gets nervous easily} &$\square$& $\square$ & $\square$  &$\square$  & $\square$
		\\\hline
		\footnotesize \textbf{feels comfortable with self} &$\square$& $\square$ & $\square$  &$\square$  & $\square$
		\\\hline
		\footnotesize \textbf{panics easily} &$\square$& $\square$ & $\square$  &$\square$  & $\square$
		\\\hline
		\footnotesize \textbf{keeps in the background} &$\square$& $\square$ & $\square$  &$\square$  & $\square$
		\\\hline
		\footnotesize \textbf{finds it difficult to get down to work} &$\square$& $\square$ & $\square$  &$\square$  & $\square$
		\\\hline
		\footnotesize \textbf{tends to vote for conservative candidates} &$\square$& $\square$ & $\square$  &$\square$  & $\square$
		\\\hline
		\footnotesize \textbf{skilled in handling social situations} &$\square$& $\square$ & $\square$  &$\square$  & $\square$
		\\\hline
		\footnotesize \textbf{wastes time} &$\square$& $\square$ & $\square$  &$\square$  & $\square$
		\\\hline
		\footnotesize \textbf{tends to be lazy} &$\square$& $\square$ & $\square$  &$\square$  & $\square$
		\\\hline
		\footnotesize \textbf{respects others} &$\square$& $\square$ & $\square$  &$\square$  & $\square$
		\\\hline
		\caption{} 
		\label{tab:longtable1}
	\end{longtable}

	\onecolumn
	\section{Commonsense Survey and Results}
	\begin{longtable}[b]{|m{4cm}||c|c|c|c|c|c|c|c|} \hline
		\footnotesize \textbf{Question} &\scriptsize RFC & \scriptsize DT & \scriptsize KNN  &\scriptsize SVM-Linear  &\scriptsize SVM-Poly  &\scriptsize SVM-rbf  &\scriptsize Perceptron  &\scriptsize MLP
		\\\hline
		{\scriptsize What are your gender/color priorities for the US commander in chief?} &\scriptsize 84\% &\scriptsize 70\% &\scriptsize 83\% &\scriptsize 77\% &\scriptsize 77\% &\scriptsize 83\% &\scriptsize 83\% &\scriptsize 72\%
		\\\hline
		{\scriptsize President Trumps travel ban is} &\scriptsize 82\% &\scriptsize 72\% &\scriptsize 78\% &\scriptsize 76\% &\scriptsize 77\% &\scriptsize 79\% &\scriptsize 70\% &\scriptsize 71\%
		\\\hline
		{\scriptsize When you're looking for work, name a way you might find out about the job opening} &\scriptsize 78\% &\scriptsize 66\% &\scriptsize 75\% &\scriptsize 79\% &\scriptsize 75\% &\scriptsize 79\% &\scriptsize 58\% &\scriptsize 70\%
		\\\hline
		{\scriptsize{An engineer who lost her father in childhood in a shooting accident hates guns and blames gunmakers for her loss.However, the only job offer that she has received is from a big gunmaker company. If she declines the job offer (refusing to make guns that can kill other kids' dads), her retired mother must sell her house to pay off her school loan.She should ...}} &\scriptsize 78\% &\scriptsize 66\% &\scriptsize 75\% &\scriptsize 79\% &\scriptsize 75\% &\scriptsize 79\% &\scriptsize 58\% &\scriptsize 70\%
		\\\hline
		{\scriptsize Name something your body tells you it's time to do.} &\scriptsize 73\% &\scriptsize 59\% &\scriptsize 70\% &\scriptsize 70\% &\scriptsize 70\% &\scriptsize 72\% &\scriptsize 47\% &\scriptsize 64\%
		\\\hline
		{\scriptsize What is your opinion about creating united health care insurance by slightly increasing the sales/income tax?} &\scriptsize 73\% &\scriptsize 59\% &\scriptsize 70\% &\scriptsize 70\% &\scriptsize 70\% &\scriptsize 72\% &\scriptsize 47\% &\scriptsize 64\%
		\\\hline
		{\scriptsize You share values with} &\scriptsize 70\% &\scriptsize 59\% &\scriptsize 54\% &\scriptsize 64\% &\scriptsize 59\% &\scriptsize 66\% &\scriptsize 47\% &\scriptsize 57\%
		\\\hline
		{\scriptsize Name something that people like to do in their backyard} &\scriptsize 70\% &\scriptsize 59\% &\scriptsize 54\% &\scriptsize 64\% &\scriptsize 59\% &\scriptsize 66\% &\scriptsize 47\% &\scriptsize 57\%
		\\\hline
		{\scriptsize Name something that's embarrassing to fall asleep while doing.} &\scriptsize 69\% &\scriptsize 61\% &\scriptsize 65\% &\scriptsize 65\% &\scriptsize 65\% &\scriptsize 69\% &\scriptsize 63\% &\scriptsize 60\%
		\\\hline
		{\scriptsize A patient who has agreed to donate his vital organs is in a coma and lives only with the assistance of life-support machines,but will pass away in the next few days anyway. There are a few patients who need the organs withing a few hours to survive. Which one do you agree with:} &\scriptsize 67\% &\scriptsize 54\% &\scriptsize 64\% &\scriptsize 65\% &\scriptsize 62\% &\scriptsize 67\% &\scriptsize 62\% &\scriptsize 68\%
		\\\hline
		{\scriptsize Name a reason people change their names} &\scriptsize 66\% &\scriptsize 54\% &\scriptsize 65\% &\scriptsize 60\% &\scriptsize 61\% &\scriptsize 66\% &\scriptsize 59\% &\scriptsize 61\%
		\\\hline
		{\scriptsize Why did Hillary Clinton lose the presidential election?} &\scriptsize 65\% &\scriptsize 59\% &\scriptsize 57\% &\scriptsize 62\% &\scriptsize 61\% &\scriptsize 65\% &\scriptsize 58\% &\scriptsize 58\%
		\\\hline
		{\scriptsize If a scientific theory does not agree with your religious belief:} &\scriptsize 65\% &\scriptsize 55\% &\scriptsize 60\% &\scriptsize 61\% &\scriptsize 65\% &\scriptsize 65\% &\scriptsize 58\% &\scriptsize 56\%
		\\\hline
		{\scriptsize Tell me the most important habit a mom should teach her son.} &\scriptsize 63\% &\scriptsize 54\% &\scriptsize 56\% &\scriptsize 64\% &\scriptsize 62\% &\scriptsize 60\% &\scriptsize 55\% &\scriptsize 60\%
		\\\hline				
		{\scriptsize Name something that people do to their faces} &\scriptsize 62\% &\scriptsize 56\% &\scriptsize 57\% &\scriptsize 62\% &\scriptsize 59\% &\scriptsize 65\% &\scriptsize 58\% &\scriptsize 62\%
		\\\hline				
		{\scriptsize If you were arrested and could make only one call, whom would you call?} &\scriptsize 62\% &\scriptsize 50\% &\scriptsize 56\% &\scriptsize 59\% &\scriptsize 58\% &\scriptsize 62\% &\scriptsize 59\% &\scriptsize 50\%
		\\\hline				
		{\scriptsize Name something you've stayed up all night worrying about.} &\scriptsize 61\% &\scriptsize 55\% &\scriptsize 56\% &\scriptsize 55\% &\scriptsize 54\% &\scriptsize 59\% &\scriptsize 53\% &\scriptsize 52\%
		\\\hline				
		{\scriptsize Is this a big deal if the fired missile from Yemen to Saudi Arabia was in fact made in Iran?} &\scriptsize 60\% &\scriptsize 54\% &\scriptsize 57\% &\scriptsize 59\% &\scriptsize 59\% &\scriptsize 61\% &\scriptsize 44\% &\scriptsize 58\%
		\\\hline				
		\scriptsize{A US-based solar panel manufacturing company  can no longer compete with fully-automated companies in Asia that manufacture solar panels 50\% cheaper. Hundreds of workers will lose their jobs in the US if the company is closed. The company should ... }&\scriptsize 60\% &\scriptsize 55\% &\scriptsize 56\% &\scriptsize 58\% &\scriptsize 60\% &\scriptsize 63\% &\scriptsize 60\% &\scriptsize 55\%
		\\\hline				
		{\scriptsize Name something that your boss could tell you that would come as a big shock.} &\scriptsize 60\% &\scriptsize 55\% &\scriptsize 54\% &\scriptsize 55\% &\scriptsize 55\% &\scriptsize 55\% &\scriptsize 52\% &\scriptsize 52\%
		\\\hline				
		{\scriptsize In a smartphone, the main issue for you is} &\scriptsize 59\% &\scriptsize 55\% &\scriptsize 57\% &\scriptsize 59\% &\scriptsize 60\% &\scriptsize 59\% &\scriptsize 50\% &\scriptsize 56\%
		\\\hline				
		{\scriptsize Name something people do when riding in the back of a taxicab} &\scriptsize 59\% &\scriptsize 53\% &\scriptsize 52\% &\scriptsize 57\% &\scriptsize 55\% &\scriptsize 57\% &\scriptsize 57\% &\scriptsize 43\%
		\\\hline
		{\scriptsize If money is not an issue, you would like to buy} &\scriptsize 59\% &\scriptsize 55\% &\scriptsize 51\% &\scriptsize 50\% &\scriptsize 54\% &\scriptsize 57\% &\scriptsize 54\% &\scriptsize 50\%
		\\\hline
		{\scriptsize Name something specific you might remember about one of your teachers from high school.} &\scriptsize 58\% &\scriptsize 52\% &\scriptsize 53\% &\scriptsize 59\% &\scriptsize 59\% &\scriptsize 59\% &\scriptsize 53\% &\scriptsize 53\%
		\\\hline
		{\scriptsize Name a reason a young man might join the army} &\scriptsize 57\% &\scriptsize 52\% &\scriptsize 51\% &\scriptsize 54\% &\scriptsize 56\% &\scriptsize 57\% &\scriptsize 50\% &\scriptsize 55\%
		\\\hline
		{\scriptsize Name a reason you might ask the manager of a hotel for another room.} &\scriptsize 57\% &\scriptsize 53\% &\scriptsize 55\% &\scriptsize 62\% &\scriptsize 60\% &\scriptsize 60\% &\scriptsize 58\% &\scriptsize 57\%
		\\\hline
		{\scriptsize You are tired of a driver who is tailgating you for over an hour, what would you do?} &\scriptsize 57\% &\scriptsize 54\% &\scriptsize 56\% &\scriptsize 56\% &\scriptsize 56\% &\scriptsize 58\% &\scriptsize 58\% &\scriptsize 55\%
		\\\hline		
		{\scriptsize Name a way you can tell that your child is lying} &\scriptsize 57\% &\scriptsize 52\% &\scriptsize 53\% &\scriptsize 45\% &\scriptsize 48\% &\scriptsize 54\% &\scriptsize 53\% &\scriptsize 50\%
		\\\hline		
		{\scriptsize If you have ever been fined for any driving violations, it has most likely been for ...} &\scriptsize 57\% &\scriptsize 48\% &\scriptsize 57\% &\scriptsize 56\% &\scriptsize 58\% &\scriptsize 57\% &\scriptsize 52\% &\scriptsize 59\%
		\\\hline		
		{\scriptsize The politician who wins the election is usually the person with the most what?} &\scriptsize 57\% &\scriptsize 52\% &\scriptsize 56\% &\scriptsize 53\% &\scriptsize 54\% &\scriptsize 56\% &\scriptsize 48\% &\scriptsize 49\%
		\\\hline		
		\scriptsize Name something that you'd like to do all day without feeling any guilt. &\scriptsize 56\% &\scriptsize 51\% &\scriptsize 54\% &\scriptsize 53\% &\scriptsize 55\% &\scriptsize 58\% &\scriptsize 56\% &\scriptsize 51\%
		\\\hline		
		\scriptsize Which food do you think should be chosen as the "National Food" of America? &\scriptsize 56\% &\scriptsize 49\% &\scriptsize 56\% &\scriptsize 55\% &\scriptsize 57\% &\scriptsize 56\% &\scriptsize 54\% &\scriptsize 52\%
		\\\hline		
		\scriptsize Name something most people clean at least once a day. &\scriptsize 55\% &\scriptsize 55\% &\scriptsize 54\% &\scriptsize 55\% &\scriptsize 52\% &\scriptsize 54\% &\scriptsize 50\% &\scriptsize 52\%
		\\\hline		
		\scriptsize Name a reason why people sometimes need to take a day off from the work. &\scriptsize 54\% &\scriptsize 51\% &\scriptsize 53\% &\scriptsize 52\% &\scriptsize 51\% &\scriptsize 58\% &\scriptsize 54\% &\scriptsize 54\%
		\\\hline		
		\scriptsize What's the hardest part of your budget to cut out? &\scriptsize 54\% &\scriptsize 39\% &\scriptsize 52\% &\scriptsize 43\% &\scriptsize 47\% &\scriptsize 54\% &\scriptsize 33\% &\scriptsize 54\%
		\\\hline		
		{\scriptsize Name a tip people give you on how to get rich.} &\scriptsize 53\% &\scriptsize 52\% &\scriptsize 52\% &\scriptsize 49\% &\scriptsize 48\% &\scriptsize 56\% &\scriptsize 49\% &\scriptsize 52\%
		\\\hline		
		{\scriptsize Where would you look to find out how old someone is?} &\scriptsize 52\% &\scriptsize 53\% &\scriptsize 51\% &\scriptsize 55\% &\scriptsize 54\% &\scriptsize 55\% &\scriptsize 49\% &\scriptsize 53\%
		\\\hline		
		{\scriptsize Name a technique people use to remember things.} &\scriptsize 51\% &\scriptsize 50\% &\scriptsize 48\% &\scriptsize 48\% &\scriptsize 50\% &\scriptsize 51\% &\scriptsize 53\% &\scriptsize 52\%
		\\\hline		
		{\scriptsize Name something that's important to parents when house hunting?} &\scriptsize 50\% &\scriptsize 39\% &\scriptsize 42\% &\scriptsize 49\% &\scriptsize 46\% &\scriptsize 50\% &\scriptsize 46\% &\scriptsize 38\%
		\\\hline		
		{\scriptsize Tell me a reason why a man would want to have a son.} &\scriptsize 50\% &\scriptsize 48\% &\scriptsize 49\% &\scriptsize 51\% &\scriptsize 49\% &\scriptsize 52\% &\scriptsize 49\% &\scriptsize 48\%
		\\\hline		
		{\scriptsize Name a reason you might make your spouse get out of bed at the middle of the night } &\scriptsize 48\% &\scriptsize 43\% &\scriptsize 37\% &\scriptsize 44\% &\scriptsize 42\% &\scriptsize 40\% &\scriptsize 39\% &\scriptsize 39\%
		\\\hline		
		{\scriptsize Many jobs will be lost in the next few decades due to smart robots and unemployment will thus increase. Therefore, } &\scriptsize 48\% &\scriptsize 36\% &\scriptsize 39\% &\scriptsize 40\% &\scriptsize 41\% &\scriptsize 48\% &\scriptsize 39\% &\scriptsize 37\%
		\\\hline		
		{\scriptsize When you pick a password for your phone, you pick a password that is.} &\scriptsize 48\% &\scriptsize 31\% &\scriptsize 41\% &\scriptsize 40\% &\scriptsize 40\% &\scriptsize 48\% &\scriptsize 37\% &\scriptsize 42\%
		\\\hline	
		\scriptsize On a sunny day, in a freeway with the maximum speed of 60 miles per hour if there is no police officer, you will typically drive with a speed ... &\scriptsize 50\% &\scriptsize 51\% &\scriptsize 52\% &\scriptsize 50\% &\scriptsize 52\% &\scriptsize 50\% &\scriptsize 48\% &\scriptsize 57\%
		\\\hline	
		\scriptsize If money is not an issue, you will buy a driver-less car &\scriptsize 47\% &\scriptsize 43\% &\scriptsize 39\% &\scriptsize 42\% &\scriptsize 44\% &\scriptsize 48\% &\scriptsize 45\% &\scriptsize 42\%
		\\\hline	
		\scriptsize What should the US do if we learn that the false nuclear alarm in Hawaii was real but the North Korean missile exploded before reaching Hawaii: &\scriptsize 47\% &\scriptsize 39\% &\scriptsize 44\% &\scriptsize 40\% &\scriptsize 41\% &\scriptsize 48\% &\scriptsize 35\% &\scriptsize 41\%
		\\\hline	
		\scriptsize If you are not in a hurry, which route would you choose to your destination: &\scriptsize 45\% &\scriptsize 37\% &\scriptsize 34\% &\scriptsize 39\% &\scriptsize 40\% &\scriptsize 47\% &\scriptsize 40\% &\scriptsize 35\%
		\\\hline	
		\scriptsize True democracy can be implemented using smartphone voting, as people can vote every day on important issues and there is no need to have a costly congress or senate anymore. The people will finally be in charge! &\scriptsize 43\% &\scriptsize 35\% &\scriptsize 36\% &\scriptsize 37\% &\scriptsize 36\% &\scriptsize 41\% &\scriptsize 38\% &\scriptsize 38\%
		\\\hline
		\scriptsize You really like a luxury car that costs \$70,000 but you can only afford \$40,000. &\scriptsize 43\% &\scriptsize 34\% &\scriptsize 37\% &\scriptsize 33\% &\scriptsize 35\% &\scriptsize 43\% &\scriptsize 43\% &\scriptsize 38\%
		\\\hline	
		\scriptsize Name something you hope you never find in your house. &\scriptsize 40\% &\scriptsize 24\% &\scriptsize 34\% &\scriptsize 30\% &\scriptsize 32\% &\scriptsize 39\% &\scriptsize 24\% &\scriptsize 29\%
		\\\hline	
		\scriptsize Name a subject that sons discuss with their fathers rather than with their mothers. &\scriptsize 40\% &\scriptsize 37\% &\scriptsize 39\% &\scriptsize 43\% &\scriptsize 39\% &\scriptsize 45\% &\scriptsize 30\% &\scriptsize 38\%
		\\\hline	
		\scriptsize Besides being fast, name something people like about fast food. &\scriptsize 39\% &\scriptsize 35\% &\scriptsize 35\% &\scriptsize 35\% &\scriptsize 36\% &\scriptsize 40\% &\scriptsize 36\% &\scriptsize 35\%
		\\\hline	
		{\scriptsize Name something one family member might steal from another.} &\scriptsize 37\% &\scriptsize 38\% &\scriptsize 30\% &\scriptsize 32\% &\scriptsize 32\% &\scriptsize 37\% &\scriptsize 34\% &\scriptsize 33\%
		\\\hline	
		{\scriptsize A study at the University of Washington showed that by increasing the minimum wage, unemployment rate will increase, therefore} &\scriptsize 37\% &\scriptsize 29\% &\scriptsize 36\% &\scriptsize 33\% &\scriptsize 34\% &\scriptsize 37\% &\scriptsize 29\% &\scriptsize 34\%
		\\\hline	
		{\scriptsize Name a reason you might pull your car over to the side of the road.} &\scriptsize 37\% &\scriptsize 36\% &\scriptsize 36\% &\scriptsize 30\% &\scriptsize 30\% &\scriptsize 37\% &\scriptsize 26\% &\scriptsize 31\%
		\\\hline	
		{\scriptsize Name something that would be smart to know how to ask for in a foreign country.} &\scriptsize 37\% &\scriptsize 35\% &\scriptsize 33\% &\scriptsize 36\% &\scriptsize 32\% &\scriptsize 34\% &\scriptsize 36\% &\scriptsize 32\%
		\\\hline	
		{\scriptsize Name something you do while driving, even though you know it's wrong} &\scriptsize 36\% &\scriptsize 32\% &\scriptsize 35\% &\scriptsize 34\% &\scriptsize 34\% &\scriptsize 40\% &\scriptsize 31\% &\scriptsize 33\%
		\\\hline	
		{\scriptsize As micro sized robots will be developed soon,people can be filmed anytime and anywhere without their knowledge, which will destroy their privacy.} &\scriptsize 35\% &\scriptsize 26\% &\scriptsize 34\% &\scriptsize 36\% &\scriptsize 36\% &\scriptsize 37\% &\scriptsize 32\% &\scriptsize 31\%
		\\\hline	
		{\scriptsize Name something people do to guard against wrinkles} &\scriptsize 34\% &\scriptsize 27\% &\scriptsize 28\% &\scriptsize 31\% &\scriptsize 31\% &\scriptsize 34\% &\scriptsize 32\% &\scriptsize 36\%
		\\\hline	
		{\scriptsize Name a reason why someone might go a whole day without eating.} &\scriptsize 33\% &\scriptsize 36\% &\scriptsize 34\% &\scriptsize 36\% &\scriptsize 31\% &\scriptsize 34\% &\scriptsize 34\% &\scriptsize 34\%
		\\\hline	
		{\scriptsize Name a Doctor's order people never take seriously.} &\scriptsize 32\% &\scriptsize 25\% &\scriptsize 27\% &\scriptsize 31\% &\scriptsize 33\% &\scriptsize 32\% &\scriptsize 29\% &\scriptsize 30\%
		\\\hline	
		{\scriptsize We asked 100 married women... If your husband asked you for a divorce Sunday night, what's the first thing you'd do Monday morning?} &\scriptsize 30\% &\scriptsize 30\% &\scriptsize 27\% &\scriptsize 28\% &\scriptsize 26\% &\scriptsize 29\% &\scriptsize 31\% &\scriptsize 31\%
		\\\hline	
		{\scriptsize Name a question a person who's afraid of flight might ask a flight attendant.} &\scriptsize 30\% &\scriptsize 24\% &\scriptsize 26\% &\scriptsize 22\% &\scriptsize 23\% &\scriptsize 32\% &\scriptsize 30\% &\scriptsize 32\%
		\\\hline	
		{\scriptsize If time flies when you're having fun, name a place it crawls} &\scriptsize 29\% &\scriptsize 24\% &\scriptsize 30\% &\scriptsize 32\% &\scriptsize 30\% &\scriptsize 30\% &\scriptsize 26\% &\scriptsize 29\%
		\\\hline	
		{\scriptsize Name something specific everyone complains about} &\scriptsize 21\% &\scriptsize 18\% &\scriptsize 22\% &\scriptsize 23\% &\scriptsize 18\% &\scriptsize 25\% &\scriptsize 16\% &\scriptsize 27\%
		\\\hline	
		\caption{} 
		\label{tab:longtable}
	\end{longtable}	
	\begin{adjustbox}{angle=90}
		\begin{tabular}{|c|c|}  \hline
			\textbf{Question} & \textbf{Answers}\\ \hline
			\small{Name a subject that sons discuss with their fathers rather than with their mothers.} & \tabitem \small Girls \\ & \tabitem \small Sports \\ & \tabitem \small{Becoming a man} \\ & \tabitem \small{Birds and bees} \\ \hline
			\small{Name a technique people use to remember things.} & \tabitem \small{Thinking about it} \\ & \tabitem \small{Repeating it} \\ & \tabitem \small{Word association} \\ & \tabitem \small{String on finger} \\ & \tabitem \small{Writing it down} \\ \hline
			
			\small{Name something people do to guard against wrinkles} & \tabitem \small{Botox} \\ & \tabitem \small{Stay out of sun} \\ & \tabitem \small{Iron clothes} \\ & \tabitem \small{Sunscreen} \\ & \tabitem \small{Facial cream} \\ & \tabitem \small{Wear makeup} \\ \hline
			\small{Name something people like to do in their backyard} & \tabitem \small Games \\ & \tabitem \small Barbecue \\ & \tabitem \small{Catch some rays} \\ & \tabitem \small{Garden/Mow grass} \\ & \tabitem \small Swim \\ \hline
			\small{Name something people do when riding in the back of a taxicab} & \tabitem \small{Talk to the cabby} \\ & \tabitem \small{Look out the window} \\ & \tabitem \small{Talk on cell} \\ & \tabitem \small{Give directions} \\ & \tabitem \small Read \\ & \tabitem \small{Watch meter} \\ & \tabitem \small Sleep \\ \hline
			\small{Name a question a person who's afraid of flight might ask a flight attendant.} & \tabitem \small{Will we crash?} \\ & \tabitem \small{You certified?} \\ & \tabitem \small{Is the pilot good?} \\ & \tabitem \small{How long is the flight?} \\ & \tabitem \small{How's the weather} \\ & \tabitem \small{Is the plane safe?} \\ \hline
			\small{We asked 100 married women...If your husband asked you for a divorce Sunday night,} & \tabitem \small{Clean out the bank} \\
			
			\small{what's the first thing you'd do Monday morning?} & \tabitem \small{Kick him out} \\ & \tabitem \small{Leave the house} \\ & \tabitem \small{Ask him why} \\  & \tabitem \small{See a lawyer/File} \\ & \tabitem \small{Change the locks} \\ \hline
			\small{When you're looking for work, name a way you might find out about the job openning} & \tabitem \small{Newspaper} \\ & \tabitem \small{Friend} \\ & \tabitem \small{Internet} \\ & \tabitem \small{Unemployment office} \\ \hline
		\end{tabular}
	\end{adjustbox}
	\begin{adjustbox}{angle=90}
		\begin{tabular}{|c|c|}  \hline

			\small{Name something you do while driving, even though you know it's wrong} & \tabitem \small{Speed} \\ & \tabitem \small{Smoke cigarette} \\ & \tabitem \small{Not use the seat belts} \\ & \tabitem \small{Eat/Drink} \\ & \tabitem \small{Don't use the signals} \\ & \tabitem \small{Talk on the phone} \\ \hline
			\small{Name a doctor's order people never take seriously} & \tabitem \small{Stop smoking} \\ & \tabitem \small{Get exercise} \\ & \tabitem \small{Plenty of rest} \\ & \tabitem \small{Weight/Eat right} \\ & \tabitem \small{Medicine/ vitamins} \\ \hline
			\small{Where would you look to find out how old someone is?} & \tabitem \small{License/ wallet} \\ & \tabitem \small{Birth certificate} \\ & \tabitem \small{Their hands} \\ & \tabitem \small{Their eyes} \\ & \tabitem \small{Ask family members} \\ & \tabitem \small{Ask them} \\ \hline		
			\small{Name something that would be smart to know how to ask for in a foreign country} & \tabitem \small{The embassy} \\ & \tabitem \small{Food} \\ & \tabitem \small{Directions} \\ & \tabitem \small{Bathroom} \\ & \tabitem \small{Help} \\ \hline
			\small{Name something that would be smart to know how to ask for in a foreign country} & \tabitem \small{The embassy} \\ & \tabitem \small{Food} \\ & \tabitem \small{Directions} \\ & \tabitem \small{Bathroom} \\ & \tabitem \small{Help} \\ \hline
			\normalsize{Name something that's important to parents when house hunting? } & \tabitem \small{Location} \\ & \tabitem \small{Backyard} \\ & \tabitem \small{Safety} \\ & \tabitem \small{School} \\ & \tabitem \small{Enough rooms} \\ \hline	
			\small{Besides being fast, name something people like about fast food.} & \tabitem \small{Don't cook/clean} \\ & \tabitem \small{Taste/flavor} \\ & \tabitem \small{Variety} \\ & \tabitem \small{Its hot} \\ & \tabitem \small{Price/ Cheap} \\ & \tabitem \small{Easy/ Convenient} \\ \hline
			\small{If you were arrested and could make only one call, whom would you call?} & \tabitem \small{bail bondsman} \\ & \tabitem \small{Sibling} \\ & \tabitem \small{friend} \\ & \tabitem \small{partner/ spouse} \\ & \tabitem \small{parent} \\ & \tabitem \small{lawyer} \\ \hline		
		\end{tabular}
	\end{adjustbox}
	
	\begin{adjustbox}{angle=90}
		\begin{tabular}{|c|c|}  \hline
			\small{If your cell phone rings while you're at church, who'd better be calling} & \tabitem \small{My sibling} \\ & \tabitem \small{My kids} \\ & \tabitem \small{The lord} \\ & \tabitem \small{The doctor} \\ & \tabitem \small{My spouse} \\ & \tabitem \small{My boss}\\ & \tabitem \small{My mother}\\ \hline
			\small{Tell me the most important habit a mom should teach her son} & \tabitem \small{Respect for women} \\ & \tabitem \small{Pick up after self} \\ & \tabitem \small{Toilet etiquette} \\ & \tabitem \small{Honesty} \\ & \tabitem \small{Manner} \\ & \tabitem \small{Cleanliness} \\ \hline
			\small{Name something that people do to their faces } & \tabitem \small{Have facial} \\ & \tabitem \small{Put on makeup} \\ & \tabitem \small{Shave} \\ & \tabitem \small{Get botox} \\ & \tabitem \small{Wash} \\ \hline	
			\normalsize{Name a reason a young man might join the army} & \tabitem \small{Drafted} \\ & \tabitem \small{Career} \\ & \tabitem \small{Education} \\ & \tabitem \small{Serve his country} \\ & \tabitem \small{Be all he can be} \\ \hline
			\normalsize{Name a reason people change their names} & \tabitem \small{Divorces} \\ & \tabitem \small{Dislike their real name} \\ & \tabitem \small{Get married} \\ & \tabitem \small{Hide} \\ \hline		
			\normalsize{ Name something you'd like to do all day without feeling any guilt} & \tabitem \small{Shopping} \\ & \tabitem \small{Watching TV/ Movies} \\ & \tabitem \small{Sleeping/Relaxing} \\ & \tabitem \small{Reading} \\ & \tabitem \small{Eating} \\ \hline
			\normalsize{Name a way you can tell that your child is lying.} & \tabitem \small{Look in the eye} \\ & \tabitem \small{Stuttering} \\ & \tabitem \small{Fidget} \\ & \tabitem \small{Turn red} \\ & \tabitem \small{Expression} \\ & \tabitem \small{No eye contact} \\ \hline
			\normalsize{Name a reason why people sometimes need to take a day off from the work}  & \tabitem \small{Stress relief} \\ & \tabitem \small{Tired} \\ & \tabitem \small{Hangover} \\ & \tabitem \small{Sick} \\ \hline			
		\end{tabular}
	\end{adjustbox}

	\begin{adjustbox}{angle=90}
		\begin{tabular}{|c|c|}  \hline
			\normalsize{Name a reason you might make your spouse get out of bed at the middle of the night} & \tabitem \small{Get a glass of water} \\ & \tabitem \small{Check the noise} \\ & \tabitem \small{Snoring} \\ & \tabitem \small{Feel sick} \\ & \tabitem \small{Attend to the baby} \\ \hline
			\normalsize{Name something your body tells you it's time to do} & \tabitem \small{Eat} \\ & \tabitem \small{Sleep} \\ & \tabitem \small{Wake up} \\ & \tabitem \small{Go to bathroom} \\ \hline		
			\normalsize{Name something that your boss could tell you that would come as a big shock. } & \tabitem \small{Promotion/raise} \\ & \tabitem \small{Buisness closing} \\ & \tabitem \small{Boss is gay} \\ & \tabitem \small{You do bad work} \\ & \tabitem \small{You're fired} \\ \hline
			\normalsize{If time flies when you're having fun, name a place it crawls} & \tabitem \small{Jail} \\ & \tabitem \small{In the traffic} \\ & \tabitem \small{DMV} \\ & \tabitem \small{Work} \\ & \tabitem \small{Church} \\ & \tabitem \small{School} \\ \hline
			\normalsize{What's the hardest part of your budget to cut out?} & \tabitem \small{Shopping/ clothes} \\ & \tabitem \small{Entertainment} \\ & \tabitem \small{Rent/Mortgage} \\ & \tabitem \small{Food/Eating out} \\ \hline	
			\normalsize{Which food do you think should be chosen as the "National Food" of America?} & \tabitem \small{Fries} \\ & \tabitem \small{Pizza} \\ & \tabitem \small{Fried chicken} \\ & \tabitem \small{Hotdog} \\ & \tabitem \small{Ice cream} \\ & \tabitem \small{Turkey} \\ & \tabitem \small{Burger} \\ \hline	
			\normalsize{Name something specific everyone complains about} & \tabitem \small{Taxes} \\ & \tabitem \small{Traffic} \\ & \tabitem \small{Weather} \\ & \tabitem \small{Money} \\ & \tabitem \small{Work} \\ \hline		
			\normalsize{Name something that's embarrassing to fall asleep while doing} & \tabitem \small{Making love} \\ & \tabitem \small{Eating} \\ & \tabitem \small{Drinking} \\ & \tabitem \small{Talking/phone} \\ & \tabitem \small{Driving} \\ \hline	
			\normalsize{Name something specific you might remember about one of your teachers from high school.} & \tabitem \small{Nice/good teacher} \\ & \tabitem \small{Good looking} \\ & \tabitem \small{Way they dressed} \\ & \tabitem \small{Their glasses} \\ & \tabitem \small{Mean/yelled} \\ \hline
		\end{tabular}
	\end{adjustbox}

	\begin{adjustbox}{angle=90}
		\begin{tabular}{|c|c|}  \hline

			\normalsize{Name something most people clean at least once a day.} & \tabitem \small{Hands} \\ & \tabitem \small{Bathroom} \\ & \tabitem \small{Dishes/Sink} \\ & \tabitem \small{Face} \\ & \tabitem \small{Teeth} \\ & \tabitem \small{Body} \\ \hline	
			\normalsize{Name something one family member might steal from another.} & \tabitem \small{Food} \\ & \tabitem \small{Money} \\ & \tabitem \small{Lover} \\ & \tabitem \small{Clothes/Shoes} \\ & \tabitem \small{Jewlery} \\ \hline	
			\normalsize{Name something specific everyone complains aboutName a reason you might ask the manager of a hotel for another room.} & \tabitem \small{Bigger bed/room} \\ & \tabitem \small{Thermostat broken} \\ & \tabitem \small{No TV/Cable} \\ & \tabitem \small{Bed/room dirty} \\ & \tabitem \small{Noisy} \\ & \tabitem \small{Need nonsmoking} \\ \hline		
			\normalsize{Name a tip people give you on how to get rich.} & \tabitem \small{Work hard} \\ & \tabitem \small{Save money} \\ & \tabitem \small{Gamble/lottery} \\ & \tabitem \small{Invest} \\ \hline
			\normalsize{Name a reason why someone might go a whole day without eating.} & \tabitem \small{Fasting/dieting} \\ & \tabitem \small{Forgot/busy} \\ & \tabitem \small{Depressed/stress} \\ & \tabitem \small{Sick} \\ \hline
			\normalsize{Name a reason you might pull you car over to the side of the road.} & \tabitem \small{Flat tire} \\ & \tabitem \small{Phone} \\ & \tabitem \small{Police} \\ & \tabitem \small{Out of gas} \\ & \tabitem \small{Ambulance/Fire} \\ \hline	
			\normalsize{Name something you hope you never find in your house. } & \tabitem \small{Snake} \\ & \tabitem \small{Insect} \\ & \tabitem \small{Dead body} \\ & \tabitem \small{Rodent} \\ & \tabitem \small{Burglar} \\ & \tabitem \small{Spouse cheating} \\ \hline	
			\normalsize{The politician who wins the election is usually the person with the most what?} & \tabitem \small{Vote} \\ & \tabitem \small{Money} \\ & \tabitem \small{Charisma} \\ \hline		
			\normalsize{Tell me a reason why a man would want to have a son.} & \tabitem \small{To play sports} \\ & \tabitem \small{To be like him} \\ & \tabitem \small{To carry on name} \\ & \tabitem \small{To help when old} \\ & \tabitem \small{To love} \\ \hline		
		\end{tabular}
	\end{adjustbox}

	\begin{adjustbox}{angle=90}
		\begin{tabular}{|c|c|}  \hline

			\normalsize{Name something that you've stayed up all night worrying about.} & \tabitem \small{Kids/ family} \\ & \tabitem \small{School/Test} \\ & \tabitem \small{Health issues} \\ & \tabitem \small{Money/ Bills} \\ & \tabitem \small{Job/interview} \\ \hline
			\small{An engineer who lost her father in childhood in a shooting accident and blames gunmakers for her loss.}  & \tabitem \small{Accept, look for other jobs} 
			\\ 
			\small{However, the only job offer that she has received is from a gunmaker company.} & \tabitem \small{Decline, ask her mother to sell the house to pay off the loan} 
			\\ 
			\small{If she declines the job offer (refusing to make guns that can kill other kids' dads),} & \tabitem \small{Deline, look for another job, pay loan with creit card} 
			\\ 
			\small{her retired mother must sell her house to pay off her school loan. She should} & \tabitem \small{Decline, look for anothe job, do not pay off the loan} \\ \hline	
			
			\small{A US-based solar panel manufacturing company can no longer compete}  & \tabitem \small{Lobby and add tarrifs to imported solar panels} 
			\\ 
			\small{ with fully-automated companies in Asia that manufacture solar panels 50\% cheaper. } & \tabitem \small{spread false info regarding child abuse labor in Asia} 
			\\ 
			\small{Hundreds of workers will lose their jobs in the US if the company is closed. The company should} & \tabitem \small{Hire low wage international workers} \\ & \tabitem \small{Be closed and workers move to better jobs} \\ \hline	
			
			\small{In a smartphone, the main issue for you is }  & \tabitem \small{Speed} \\  & \tabitem \small{Memory size}\\ & \tabitem \small{Display quality and size} \\ & \tabitem \small{Battery's lifespan} \\ \hline	
			\small{Through a small increase in sales/income tax, a universal healthcare insurance plan}  & \tabitem \small{No Universal healthcare isn't good for high quality medical services} 
			\\ 
			\small{ can be created in the US. Do you like this idea?} & \tabitem \small{Yes, this is a great idea!} \\ & \tabitem \small{No univesal healthcare is good, but the taxes are already high} \\ & \tabitem \small{No, Univesal healthcare is not good} \\ & \small{religous people shouldn't pay to fund abortion} \\ & \small{medical bills for a drunk driver}
			\\ \hline	
			
			\small{You really like a luxury car that costs \$70,000, but you can only afford \$40,000.}  & \tabitem \small{you lease cars untill you can buy the luxury car} \\  & \tabitem \small{You buy the best new car that you can afford for \$40,000}\\ & \tabitem \small{You just buy the car you need, for less than \$40,000} \\ & \tabitem \small{You buy a used model of the luxury car for \$40,000.} \\ \hline
			
			\small{When you pick a password for your phone, you pick one thar is}  & \tabitem \small{Easy to remember and keep it unchanged} \\  & \tabitem \small{Difficult to remember and keep it unchanged}\\ & \tabitem \small{Difficult to remember and change it regularly} \\ & \tabitem \small{Easy to remember and you change it regularly} \\ \hline			
			
			\small{You share values with}  & \tabitem \small{Democratic party} \\  & \tabitem \small{Republican party}\\ & \tabitem \small{Independent/Green party} \\ & \tabitem \small{None} \\ \hline
			
			\small{If money is not an issue, you would like to buy}  & \tabitem \small{Samsung smartphones} \\  & \tabitem \small{Apple iphone}\\ & \tabitem \small{HTC smartphones} \\ & \tabitem \small{LG smartphones} \\ \hline		
			
			\small{If not in a hurry, which rout you choose?}  & \tabitem \small{with least traffic} \\  & \tabitem \small{Shortest/fastest}\\ & \tabitem \small{most scenic} \\ & \tabitem \small{best/safest} \\ \hline

		\end{tabular}
	\end{adjustbox}
	
	\begin{adjustbox}{angle=90}
		\begin{tabular}{|c|c|}  \hline
			
			\small{A driver is tailgating you for over an hour, what would you do?}  & \tabitem \small{ignore it and remain patient} \\  & \tabitem \small{Speed up and leave it behind}\\ & \tabitem \small{slow down and annoy the driver} \\ & \tabitem \small{Pull over and let it pass} \\ \hline		
			
			\small{Why did Hillary Clinton lose the presidential election? }  & \tabitem \small{Russian cyberattacks and fake news} \\  & \tabitem \small{US electoral system}\\ & \tabitem \small{Voters didn't like her} \\ & \tabitem \small{FBI announcement regardin new emails on a laptop} \\ \hline	
			
			\small{What should the US do if we learn that the false nuclear alarm in Hawaii was real}  & \tabitem \small{Do nothing as no harm was caused} \\ 
			
			\small{but the North Korean missile exploded before reaching Hawaii:} &\tabitem \small{Detroy North Korea by a nuclear attack} \\ & \tabitem \small{Assassinate North Korean leader} \\ & \tabitem \small{Add more sanctions} \\ \hline	
			
			\small{What are your gender/color priorities for the US commander in chief?}  & \tabitem \small{Color an gender do not matter} \\  & \tabitem \small{Color matters but gender does not}\\ & \tabitem \small{Gender matters but color does not} \\ & \tabitem \small{Gender and color matter} \\ \hline
			
			\small{If money is not an issue, you will buy a driver-less car }  & \tabitem \small{Right away} \\  & \tabitem \small{Only when 25\% of cars are driver-less.}\\ & \tabitem \small{Only when 50\% of cars are driver-less} \\ & \tabitem \small{Only when you have no other option} \\ \hline			
			
			\small{President Trump’s travel ban is}  & \tabitem \small{Great and improves the safety of legal and patriotic US citizens} \\  & \tabitem \small{Not good, because it is discriminatory.} \\ & \tabitem \small{Not good, since not all Muslim people in the banned countries} \\ & \small{should be punished because of the Sept. 11 attack} \\ & \tabitem \small{Good, but should include more countries} \\ \hline
			
			\small{ Is this a big deal if the fired missile from Yemen to Saudi Arabia} & \tabitem \small{No, missiles could have been sent before the sanctions}\\
			
			\small{was in fact made in Iran?}  & \tabitem \small{No, Saudi Arabia uses US built missiles} \\  & \tabitem \small{Yes, it shows violation in sanctions}\\ & \tabitem \small{Yes, it shows Iran helped Yeme to attack Saudi Arabia} \\ \hline		
			
			\small{If a scientific theory does not agree with your religious belief}  & \tabitem \small{You deny the scientific theory, as science is not flawless.} \\  & \tabitem \small{You do not practice any religion.}\\ & \tabitem \small{You convert to a new religion that agrees with that theory.} \\ & \tabitem \small{You practice your religion but live with the differences.} \\ \hline	
			
			\small{A study at the University of Washington showed that by increasing the minimum
				wage}  & \tabitem \small{The study is flawed} \\ 
			
			\small{,unemployment rate will increase, therefore} &\tabitem \small{The minimum wage should be further increased to reduce unemployement} \\ & \tabitem \small{The minimum wage should be reduced} \\ & \tabitem \small{There should be income tax to support unemployed people.} \\ \hline
			
			\small{Many jobs will be lost in the next few decades due to smart robots}  & \tabitem \small{No idea!} \\
			
			\small{ and unemployment will thus increase. Therefore, } & \tabitem \small{Capitalism will collapse, since the middle class will vanish}\\ & \tabitem \small{Smart robots should be banned to save jobs.} \\ & \tabitem \small{The government should provide a ``living wage'' income} \\ & \small{for all unemployed workers.} \\ \hline	
		\end{tabular}
	\end{adjustbox}
	
	\begin{adjustbox}{angle=90}
		\begin{tabular}{|c|c|}  \hline	
			
			\small{As micro sized robots will be developed soon, people can be filmed anytime } & \tabitem \small{People will learn to protect their privacy.}\\
			
			\small{and anywhere without their knowledge, which will destroy their privacy}  & \tabitem \small{People will stop micro-sized robots.} \\  & \tabitem \small{The meaning of privacy will change}\\ & \tabitem \small{People will tolerate the loss of privacy.} \\ \hline		
			
			\small{True democracy can be implemented using smartphone voting,}  & \tabitem \small{There will be no real difference. Democracy is not the answer.} \\
			
			\small{as people can vote every day on important issues,} & \tabitem \small{You prefer a combination of smartphone voting and traditional representatives}\\ 
			
			\small{ and there is no need to have a costly congress or senate anymore. } & \tabitem \small{You prefer traditional representatives that you can trust and votes} \\
			
			\small{The people will finally be in charge!} & \tabitem \small{This is a great idea. Why should we pay to keep a broken system? } \\ \hline

		\end{tabular}
	\end{adjustbox}
	
\end{appendices}

\end{document}